\documentclass[twocolumn,amsmath,superscriptaddress,amssymb,nopacs,nofootinbib,prb]{revtex4-2}

\usepackage[usenames,dvips]{color}
\usepackage{graphicx}
\usepackage{dcolumn}
\usepackage{bm}
\usepackage{mathtools}
\usepackage{epstopdf}
\usepackage{esint}
\usepackage{xcolor}
\usepackage{dsfont}
\usepackage[outline]{contour}
\usepackage{nicefrac}

\usepackage[colorlinks=true,citecolor=magenta,linkcolor=magenta, urlcolor=magenta]{hyperref}

\newcommand{\bea}{\begin{eqnarray}}
\newcommand{\eea}{\end{eqnarray}}
\newcommand{\bt}{\textbf}

\newcommand{\phd}{\phantom{\dag}}
\newcommand{\ph}{\phantom{.}}

\newcommand{\noi}{\noindent}
\newcommand{\no}{\nonumber}

\begin{document}
\def\v#1{{\bf #1}}

\title{Anatomy of Spin and Current Generation from Magnetization Gradients\\ in Topological Insulators and Rashba Metals}

\author{Panagiotis Kotetes}
\email{kotetes@itp.ac.cn}
\affiliation{CAS Key Laboratory of Theoretical Physics, Institute of Theoretical Physics, Chinese Academy of Sciences, Beijing 100190, China}

\author{Hano O. M. Sura}
\affiliation{Niels Bohr Institute, University of Copenhagen, 2100 Copenhagen, Denmark}

\author{Brian M. Andersen}
\affiliation{Niels Bohr Institute, University of Copenhagen, 2100 Copenhagen, Denmark}

\vskip 1cm

\begin{abstract}
We explore the spin {\color{black}density} and charge currents arising on the surface of a topological insulator and in a 2D Rashba metal due to magnetization gradients. For topological insulators a single interconversion coefficient controls the generation of both quantities. This coefficient is quantized to a value proportional to the vorticity of the Dirac point {\color{black}which constitutes a hallmark of parity anomaly at finite density. As such, it also} unveils a robust route to disentangle and detect the protected states of a topological insulator on a given surface. {\color{black}In stark contrast, Rashba metals do not exhibit such anomalies since they contain an even number of helical branches. Nonetheless, also these are governed by quantized responses which, however, are not protected against weak disorder. Furthermore, we find that for Rashba metals the interconversion coefficients demonstrate discontinuities} and a nontrivial interplay upon varying the chemical potential, the strength of the spin-orbit coupling, and a pairing gap. Our results have implications for the binding between magnetic skyrmions and super\-con\-ducting vortices, the emergence of Majorana zero modes, and pave the way for supercon\-ducting diode effects mediated by out-of-plane magnetization gradients.
\end{abstract}
 
\maketitle

\section{Introduction}

The breaking of parity (${\cal P}$) and time-reversal (${\cal T}$) symmetries unlocks a plethora of unconventional phe\-no\-me\-na~\cite{Fradkin,VolovikBook}. The inherent vio\-la\-tion of pa\-ri\-ty is asso\-cia\-ted with the generation of a nonzero electric po\-la\-ri\-za\-tion $\bm{P}$~\cite{Winkler,QiTFT}. Upon its pre\-sen\-ce, a spin density $\bm{{\cal S}}\propto\bm{P}\times\bm{J}$ can be induced by injecting an electric charge current $\bm{J}$ into the system. This relation is also key for the so-called Edelstein effect~\cite{Edelstein1990}, which constitutes one of the most well-established charge-to-spin interconversion mechanisms. Reciprocity also allows for an inverse magnetoelectric effect where, due to a nonzero $\bm{P}$, the exchange (Zeeman) coupling of the electrons to a homogeneous magnetization $\bm{M}$ (magnetic field $\bm{B}$) leads to a charge current $\bm{J}$. Besides spintronics~\cite{SHERMP}, ${\cal P}$-${\cal T}$ violation is also responsible for va\-rious anomalies and fractio\-na\-li\-za\-tion effects in topological insulators (TIs)~\cite{QiHughesZhang,Ion,Nomura,QAHI,HasanKane,QiZhang}. An equally rich phenomenology emerges in superconductors (SCs) experiencing ${\cal P}$-${\cal T}$ violation~\cite{VolovikBook,Linder}. For example, charge transport in noncentrosymmetric SCs~\cite{Bulaevski,EdelsteinNCS,SigristUeda,Gorkov,Banerjee} in the presence of a magnetic field $\bm{B}$ is nonreciprocal~\cite{Rikken,Wakatsuki,TokuraNagaosa}. This is due to additional current contributions, with the simplest being proportional to $\bm{P}\times\bm{B}$~\cite{EdelsteinSC,Samokhin,Law}. The $\bm{B}$-dependence of the critical current forms the basis for the so-called superconducting diode effect, which was recently expe\-ri\-men\-tal\-ly detected in various SCs~\cite{Ando,Baugartner,Pal}.

In the majority of the above situations $\bm{B}$ or $\bm{M}$ are homogeneous, while the violation of ${\cal P}$ symmetry ty\-pi\-cal\-ly manifests itself in the presence of an odd-under-inversion spin-orbit coupling (SOC). Notably, however, inhomogeneous magnetism is capable of simultaneously vio\-la\-ting both ${\cal P}$ and ${\cal T}$ symmetries~\cite{Mostovoy,BrauneckerSOC}, without re\-qui\-ring any kind of SOC. This two-in-one effect appears particularly useful for engi\-nee\-ring topological SCs in 1D~\cite{Choy,KarstenNoSOC,Ivar,KlinovajaGraphene,NadgPerge,KotetesClassi,Selftuned,Pientka,Ojanen,Mohanta} and 2D~\cite{KotetesClassi,Nakosai,Mendler,WeiChen,MorrSkyrmions,MorrTripleQ,SteffensenPRR,KovalevRev,HuangKotetes}. Coupling (un)conventional SCs to pe\-rio\-di\-cal\-ly repeating magnetic textures for\-ming cry\-stals {\color{black} is known to lead} to a rich va\-rie\-ty of topological phases~\cite{SteffensenPRR} and nonstandard transport effects~\cite{Yokohama,Hals2,Christensen}. On the other hand, {\color{black}when isolated magnetic textures such as magnetic skyrmions are superimposed on conventional SCs, they can trap superconducting vortices~\cite{Hals,Baumard,EreminSkV,Menezes,Burmistrov,Neto} and Majorana zero modes~\cite{KlinovajaSkyrmion,Kovalev,Cren,Gornyi,Garnier,Wu,Pathak}. Evenmore, such textured magnetic defects are also predicted to} induce Yu-Shiba-Rusinov states~\cite{BalatskySkyrmion,Poyhonen}, circulating currents~\cite{Pershoguba,Bjornson,Malshukov,Flatte}, and Friedel oscillations~\cite{ZhangImp,Biswas} in SCs. The above effects stem from an inhomogeneous $\bm{M}$ and thus can be viewed as Edelstein effects induced by magnetization gra\-dients.

In this {\color{black}Manuscript}, motivated by the wide range of applications of inhomogeneous magnetism, we investigate interconversion effects mediated by magnetization gradients on (non)superconducting TI surfaces and in 2D Rashba metals. {\color{black}We are primarily interested in these two classes of experimenally accessible Rashba-type systems because they constitute two of the most prominent candidates for a vast range of applications in nanoelectronics.}

We first study the induced electric current and spin densities due to magnetization gradients in the absence of superconductivity. We reveal that in the case of a TI surface with a pristine Dirac cone, the respective zero-temperature interconversion coefficient is {\color{black}quantized since it constitutes} a to\-po\-lo\-gi\-cal invariant. In particular, it is proportional to the vor\-ti\-ci\-ty of the topologically-protected touching point of the helical surface energy di\-spersion. Hence, inferring this coefficient in experiments can be harnessed to detect the protected states on a given TI surface, which is otherwise challen\-ging to achieve with Hall transport~\cite{QiHughesZhang,HasanKane,QiZhang}. {\color{black}On the other hand,} adding a pai\-ring gap {\color{black}always} spoils the quantization of the interconversion coefficient due to nonuniversal Cooper pair contributions, which tend to suppress the effect. {\color{black}Quite remar\-ka\-bly}, we find that the coefficient ``jumps'' by a quantized value even by switching on an infinitesimally weak pairing gap, as a result of quantum anomalies~\cite{Fradkin,VolovikBook}.

The topological nature of the TI interconversion coefficient allows us to make direct predictions for a 2D Rashba metal. {\color{black} This is because, in certain limits, the latter can be effectively described as a collection of two decoupled TI helical Dirac surface states. Since systems with an even number of helical Dirac electrons are anomaly free~\cite{Fradkin,VolovikBook}, this distinction underlines the importance of comparing the TI results to those obtained for Rashba metals. In fact, our analysis immediately reveals an important difference compared to TIs, that is, in Rashba metals the coefficients for spin and current generation are generally distinct. Moreover, by restricting to the quasiclassical regime, we find that they are both zero in the absence of supercon\-duc\-ti\-vi\-ty. However, they both get switched on even when an infinitesimally weak pai\-ring gap is introduced. Notably, the coefficient associated with current generation shows a discontinuous jump equal to a universal value across this transition. The latter coefficient is also here sensitive to disor\-der~\cite{SHERMP}, which is a situation reminiscent of the intrinsic spin Hall effect~\cite{SHE}.

The remainder of our manuscript is organized in the following fashion. Section~\ref{sec:SymmetryClass} provides a symmetry based phenomenological description of the magnetoelectric effects which become accessible in a system with Rashba-type SOC when magnetization gradients are present. In Sec.~\ref{sec:MicroscopicTheory} we present the formalism for deriving the coefficients which control the various interconversion phenomena from a microscopic electronic Hamiltonian. Section~\ref{Sec:TI} discusses the arising effects on the surface of a 3D TI. There, we show that spin and current generation are both controlled by the same interconversion coefficient, which becomes quantized. In the same section, we demonstrate the topological origin behind this macroscopic quantization and investigate its stability against deviations away from the pristine Dirac cone picture, Zeeman corrections, and the influence of a pairing gap. In Sec.~\ref{sec:Rashba} we proceed by examining the case of a 2D Rashba metal, which is not expected to exhibit the anomalous effects encountered in the TI case. Nevertheless, quantization phenomena of a different origin appear also here. Moreover, the interconversion coefficients is substantially affected by the presence of a pairing gap and the Zeeman effect. Potential applications are discussed in Sec.~\ref{sec:Experimental}, while we summarize our findings in Sec.~\ref{sec:summary}. Finally, details on our calculations are given in Appendices~\ref{app:Appendix1}-\ref{app:Appendix4}.}

\section{Insights from symmetry}\label{sec:SymmetryClass}

To expose the key aspects of our work, we first {\color{black}rely on the predictive power of a symmetry analysis, in order to highlight the key mechanisms which underly the spin and current generation. Our symmetry analysis is carried out in terms of} the in-plane components $A_{x,y}(\bm{r})$ of the vector potential and all the components $M_{x,y,z}(\bm{r})$ of the magnetization. These are defined in the 2D coordinate space $\bm{r}=(x,y)$. In the remainder, we restrict our study to layered systems with C$_{4v}\times{\cal T}$ symmetry, where C$_{4v}$ is the te\-tra\-go\-nal point group. {\color{black} We remark that this choice does not restrict the generality of our approach, which can be also extended to other point groups.}

\subsection{Symmetry Classification}

{\color{black} The backbone of our symmetry classification program relies on the identification of the transformation properties of the various fields under the operations of the ensuing point group. For the C$_{4v}$ point group of interest here, we find that $(A_x(\bm{r}),A_y(\bm{r}))$ and $(M_y(\bm{r}),-M_x(\bm{r}))$ transform according to the same 2D irreducible representation of C$_{4v}$.} Therefore, the term $A_x(\bm{r})M_y(\bm{r})-A_y(\bm{r})M_x(\bm{r})$ belongs to the trivial irreducible representation {\color{black}of the group C$_{4v}\times{\cal T}$. In fact, the above symmetry invariant term is responsible for the standard Edelstein effect~\cite{Edelstein1990}.}

One observes {\color{black}similarities in the structure of the symmetry invariant term $A_x(\bm{r})M_y(\bm{r})-A_y(\bm{r})M_x(\bm{r})$} and the Rashba SOC term $\hat{p}_x\sigma_y-\hat{p}_y\sigma_x$. Indeed, the construction of the two terms relies on the symmetry equivalences $(A_x,A_y)\sim(\hat{p}_x,\hat{p}_y)$ and $(M_x,M_y)\sim(\sigma_x,\sigma_y)$. Notably, {\color{black}also the two component vector} $(\partial_yM_z(\bm{r}),-\partial_xM_z(\bm{r}))$ belongs to the same representation as $(A_x(\bm{r}),A_y(\bm{r}))$ does, thus enabling a number of {\color{black}new interconversion phenomena that we bring forward in this work.}\\

\subsection{Phenomenological Energy Density and Equations of Motion}

{\color{black}Throughout this work, the effects of interest mainly concern} the generation of spin density $\bm{{\cal S}}(\bm{r})$ and electrical current $\bm{J}(\bm{r})$, which are here viewed as conjugate fields of $\bm{M}(\bm{r})$ and $\bm{A}(\bm{r})$. The {\color{black}quantities $\bm{{\cal S}}(\bm{r})=\big<\hat{\bm{S}}(\bm{r})\big>$ and $\bm{J}(\bm{r})=\big<\hat{\bm{J}}(\bm{r})\big>$ are defined as the expectation va\-lues of the respective microscopic operators $\hat{\bm{{\cal O}}}(\bm{r})=\bm{\Psi}^\dag(\bm{r})\hat{\bm{{\cal O}}}\bm{\Psi}(\bm{r})$ where $\bm{\Psi}(\bm{r})$ denotes the electronic spinor.}

The source fields enter in the Hamiltonian through the term $-\frac{1}{2}\int d\bm{r}\ph\bm{\Psi}^\dag(\bm{r})\big[\bm{A}(\bm{r})\cdot\hat{\bm{J}}(\bm{r})+\bm{M}(\bm{r})\cdot\bm{\sigma}\big]\bm{\Psi}(\bm{r})$, where $\bm{\sigma}$ are the spin Pauli matrices and $\hat{\bm{J}}(\bm{r})$ is the respective current operator represented in coordinate space. {\color{black}More details on the microscopic approach and evaluation of the various coefficients are provided in Sec.~\ref{sec:MicroscopicTheory} and Appendices~\ref{app:Appendix1}-\ref{app:Appendix4}.}

At zero temperature the equilibrium spin and current densities are obtained from the functional derivatives:
\begin{align}
\bm{{\cal S}}(\bm{r})=-\frac{\delta E(\bm{r})}{\delta \bm{M}(\bm{r})}\quad{\rm and}\quad
\bm{J}(\bm{r})=-\frac{\delta E(\bm{r})}{\delta\bm{A}(\bm{r})}
\end{align}

\noi of the energy density $E(\bm{r})$. According to our {\color{black}symmetry} analysis, the latter obtains the form:
\begin{align}
E(\bm{r})=\left[M_z(\bm{r})-\frac{g\mu_B B_z(\bm{r})}{2}\right]\big[\chi B_z(\bm{r})+g_{\rm soc}\bm{\nabla}\cdot\bm{M}(\bm{r})\big]\no\\
-\chi_\perp^{\rm spin}\frac{\big[M_z(\bm{r})-g\mu_BB_z(\bm{r})/2\big]^2}{2}-\chi_{||}^{\rm spin}\sum_{a=x,y}\frac{M_a^2(\bm{r})}{2}\label{eq:EnergyDensity}
\end{align}

\noi where we also included the Zeeman coupling to the electrons, which enters by shifting the out-of-plane magnetization {\color{black}according to $M_z(\bm{r})\mapsto M_z(\bm{r})-g\mu_BB_z(\bm{r})/2$. Here,} $g$ denotes the gyromagnetic Land\'e factor, $\mu_B$ the Bohr magneton, and $\chi_{\perp,||}^{\rm spin}$ define the out-of- and in-plane spin susceptibi\-li\-ties of the Rashba system{\color{black}, respectively.

A few comments are in place regarding the energy density above. First of all, we remark that it is not crucial to shift the in-plane magnetization components in a si\-mi\-lar fashion to $M_z$, since the orbital coupling to the in-plane magnetic field components $B_{x,y}$ is considered to be identically zero due to the planar nature of the system. Hence, Zeeman effects due to $B_{x,y}$ can be fully absorbed in $M_{x,y}$. Further, it is also important to emphasize that the energy density of Eq.~\eqref{eq:EnergyDensity} discards the standard Edelstein term $A_x(\bm{r})M_y(\bm{r})-A_y(\bm{r})M_x(\bm{r})$ which is also ge\-ne\-ral\-ly permitted. This is only because here we restrict to responses solely emerging from magnetization gradients.

Variation of the energy density with respect to the source fields} reveals two categories of reciprocal interconversion relations:
\begin{align}
{\cal S}_z(\bm{r})=-{\cal X}B_z(\bm{r})\ph\leftrightarrow\ph
\ph\bm{J}(\bm{r})=-{\cal X}\bm{\nabla}\times\hat{\bm{z}}M_z(\bm{r}),\label{eq:CurrentDensity}\\
\bm{{\cal S}}(\bm{r})=g_{\rm soc}\bm{\nabla}M_z(\bm{r})\ph\leftrightarrow\ph{\cal S}_z(\bm{r})=-g_{\rm soc}\bm{\nabla}\cdot\bm{M}(\bm{r}).
\label{eq:SpinDensity}
\end{align}

\noi In the above, $\hat{\bm{z}}$ stands for the unit vector in the out-of-plane $z$ direction, while the quantities $\bm{r}$, $\bm{A}$, $\bm{J}$, and $\bm{\nabla}$ are understood as 2D vectors defined in the $xy$ plane. We need to remark that when the system is supercon\-ducting, the current of Eq.~\eqref{eq:CurrentDensity} is observable only as long as the London penetration depth of the SC is sufficiently long to render the Meissner screening ineffective in the region where the magnetization varies spatially.

{\color{black}We observe that the current interconversion phe\-no\-me\-na are controlled by the coefficients:
\begin{align}
{\cal X}=\chi+\chi_Z\quad{\rm and}\quad \chi_Z=\frac{g\mu_B\chi_\perp^{\rm spin}}{2}\,.
\end{align}

\noi Notably, the coefficient ${\cal X}$ arises from both the Rashba and Zeeman effects with contributions $\chi$ and $\chi_Z$, respectively, and relates magnetization and magnetic field. While more quantitative details regarding the magnitude of the Zeeman contribution to the coefficient ${\cal X}$ are discussed in Secs.~\ref{sec:ZeemanEffects} and~\ref{sec:Rashba}, here, we wish to remark that for a material with a sufficiently strong Rashba SOC the electron spin is expected to be predominantly confined in the plane.} Under this condition, $\chi_\perp^{\rm spin}$ is suppressed and, in turn, also renders $\chi_Z$ negligible compared to $\chi$, which is the primary case of interest.
Finally, note that Eq.~\eqref{eq:SpinDensity} stems from the symmetry equi\-va\-len\-ces $(M_x,M_y,M_z)\sim(A_y,-A_x,B_z)$, and dictates the nonstandard interconversion between an in-plane magnetization (spin density) and an out-of-plane spin density (magnetization) which are governed by the coefficient $g_{\rm soc}$.

\subsection{Implications for Magnetic Impurities}

The above analysis applies to inhomogeneous magnetism stemming from a variety of sources which are effectively described by a classical magnetization field $\bm{M}(\bm{r})$. For a magnetization polarized along the $z$ spin axis with a ro\-ta\-tio\-nal\-ly symmetric spatial profile $M_z(\bm{r})=M_z(|\bm{r}|)$, Eqs.~\eqref{eq:CurrentDensity} and~\eqref{eq:SpinDensity} lead to the following type of cir\-cu\-la\-ting currents and in-plane spin density:
\begin{align}
\bm{J}(\rho)={\cal X}\frac{dM_z(\rho)}{d\rho}\hat{\bm{\theta}}\quad{\rm and}\quad
\bm{{\cal S}}(\bm{r})=-g_{{\color{black}\rm soc}}\frac{dM_z(\rho)}{d\rho}\hat{\bm{\rho}},
\end{align}

\noi where $(\rho,\theta)$ are the polar coordinates $\rho=\sqrt{x^2+y^2}$ and $\tan\theta=y/x$. $\hat{\bm{\rho}}$ and $\hat{\bm{\theta}}$ define the respective unit vectors.

\begin{figure}[t!]
\begin{center}
\includegraphics[width=0.8\columnwidth]{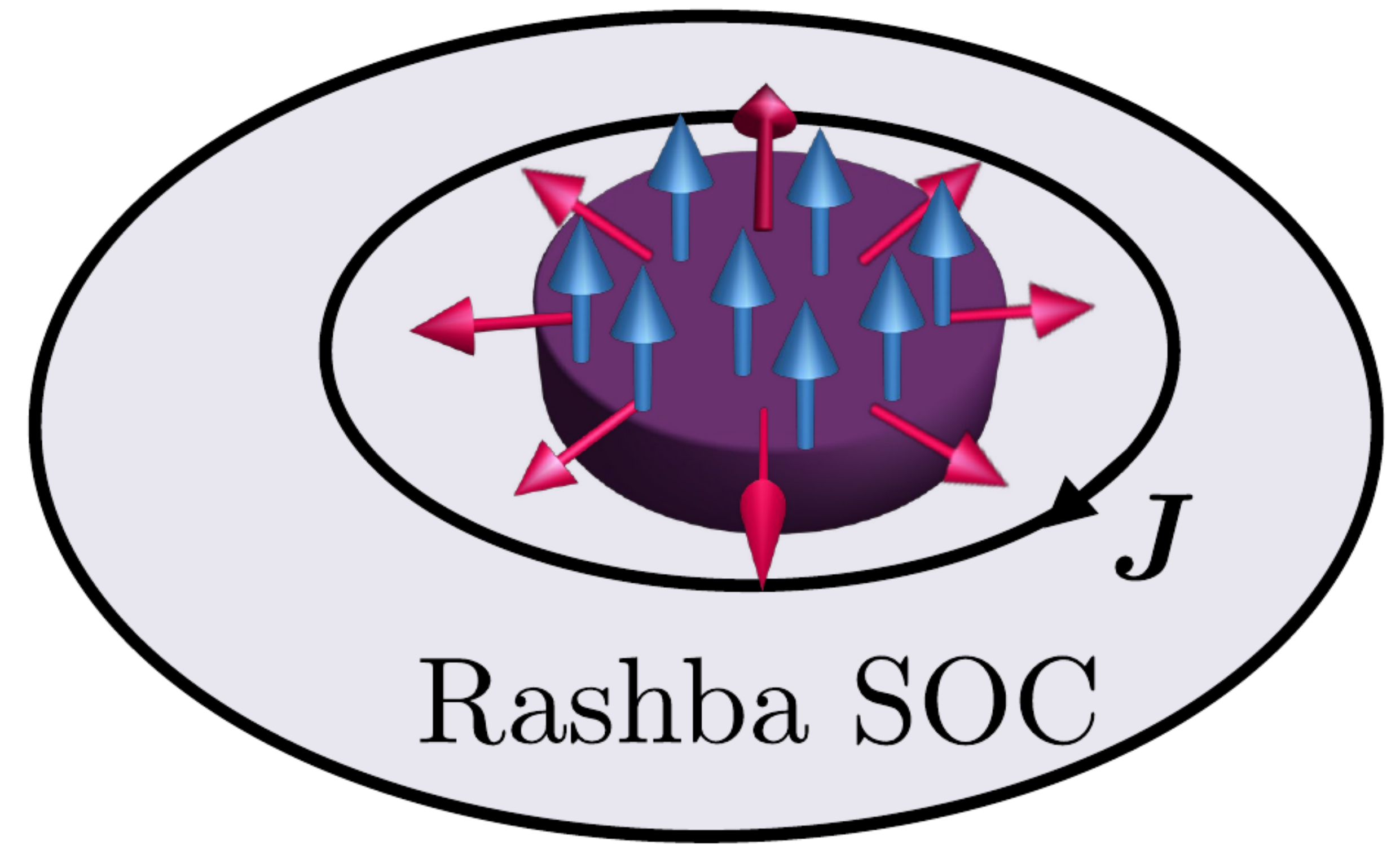}
\end{center}
\caption{{\color{black}Cartoon of an extended magnetic impurity embedded in a quasi-2D system dictated by a Rashba-type SOC. The magnetic moment of the impurity ``island" depicted with blue arrows is here assumed to be polarized out of the plane of the Rashba SOC host. Near its boundary, the magnetic impurity induces circulating electric currents accompanied by a radially-oriented in-plane magnetization which is shown with magenta arrows.}}
\label{fig:Figure1}
\end{figure}

The currents circulate about the impurity and ge\-ne\-ra\-te a nonzero vorti\-ci\-ty. Instead, the in-plane magnetic moments point along the radial direction and extend uniformly along the circumference, thus giving rise to a profile which is topologically equivalent to a magnetic vortex. {\color{black}See Fig.~\ref{fig:Figure1} for a sketch.} Our symmetry approach recovers the profiles of the current~\cite{Pershoguba,Bjornson,Flatte} and spin densities~\cite{ZhangImp,Biswas} obtained previously. However, in contrast to those works, which primarily relied on numerical methods, here we pursue exact analytical expressions for the interconversion coefficients, relevant for systems and parameter regimes that have remained so far unexplored.

\section{Microscopic Formulation of Linear Response}\label{sec:MicroscopicTheory}

{\color{black}We now proceed by focusing on the responses of concrete Rashba SOC material systems. Specifically, we consider homogeneous 2D Rashba SCs under the influence of the in-plane components $A_{x,y}(\bm{r})$ of the vector potential and all the components $M_{x,y,z}(\bm{r})$ of the magnetization. Such systems are here modeled using the Hamiltonian:}
\begin{align}
{\color{black}{\cal H}}=\frac{1}{2}\int d\bm{r}\left\{\bm{\Psi}^\dag(\bm{r})\Big[\hat{{\color{black}{\cal H}}}_0(\hat{\bm{\pi}})-\bm{M}(\bm{r})\cdot\bm{\sigma}\Big]\bm{\Psi}(\bm{r})\right\},
\label{eq:Hamiltonian}
\end{align}

\noi with the spinor $\bm{\Psi}^{\dag}(\bm{r})=\big(\psi_\uparrow^\dag(\bm{r}),\,\psi_\downarrow^\dag(\bm{r}),\,\psi_\downarrow(\bm{r}),\,-\psi_\uparrow(\bm{r})\big)$. $\psi_\sigma(\bm{r})/\psi_\sigma^\dag(\bm{r})$ annihilates/creates an electron at position $\bm{r}$ with spin projection $\sigma=\uparrow,\downarrow$. We also introduced the gauge invariant momentum $\hat{\bm{\pi}}=\hat{\bm{p}}+e\tau_z\bm{A}(\bm{r})$, where $e>0$ denotes the electric charge unit, $\hat{\bm{p}}=-i\hbar\bm{\nabla}$ the momentum operator, and $\hbar$ the reduced Planck constant. Equation~\eqref{eq:Hamiltonian} is expressed in terms of the bare Hamiltonian:
\begin{align}
\hat{\cal H}_0(\hat{\bm{p}})=\tau_z\left[\frac{\hat{\bm{p}}^2}{2m}-\mu+\upsilon\big(\hat{p}_x\sigma_y-\hat{p}_y\sigma_x\big)\right]+\Delta\tau_x.\label{eq:H0}
\end{align}

\noi Any Hamiltonian matrix, such as the above, is expressed with Kronecker pro\-ducts which are constructed using the Pauli matrices $\bm{\tau}$ and $\bm{\sigma}$, along with their respective unit matrices $\mathds{1}_{\tau,\sigma}$. These are defined in Nambu and spin spaces, respectively. Note that we omit writing the Kronecker product symbol $\otimes$ and unit matrices throughout. 

In Eq.~\eqref{eq:H0}, $\upsilon>0$ is the strength of the Rashba SOC, $m>0$ the effective mass, and $\mu$ the chemical potential. The pairing gap $\Delta\geq0$ is treated non-selfconsistently, i.e., we neglect the possible feedback effects from the magnetization on $\Delta$. This also justifies why we have not included in $\hat{\bm{\pi}}$ the contribution of the superconducting phase $\phi(\bm{r})$, which would enter by means of the miminal coupling substitution $\bm{A}(\bm{r})\mapsto\bm{A}(\bm{r})+\hbar\bm{\nabla}\phi(\bm{r})/2e$.

Our model allows us to discuss in a unified manner a Rashba metal and the surface of a TI. To obtain the response coefficients $\chi$ and $g_{\rm soc}$ in either case, we carry out a perturbative ana\-ly\-sis in terms of $\bm{A}(\bm{r})$ and $\bm{M}(\bm{r})$, which is described in detail in Appendices~\ref{app:Appendix1}-\ref{app:Appendix4}. Expan\-ding the ener\-gy density of the system $E(\bm{r})$ yields that the coefficients are obtained by evaluating the expressions:
\bea
g_{\rm soc}&=&\frac{1}{2i}\int\frac{d^3K}{(2\pi)^3}\frac{1}{2}{\rm Tr}\left[\hat{{\cal G}}_0(K)\sigma_z\frac{\partial \hat{{\cal G}}_0(K)}{\partial k_a}\sigma_a\right],\qquad
\label{eq:gspin}\\
g_{\rm orb}&=&\frac{1}{2i}\int\frac{d^3K}{(2\pi)^3}\frac{1}{2}{\rm Tr}\left[\hat{{\cal G}}_0(K)\sigma_z\varepsilon_{zab}\frac{k_a}{k_F}\frac{\partial\hat{{\cal G}}_0(K)}{\partial k_b}\right]\label{eq:gorb},\qquad\\
\chi&=&\chi_{\rm orb}+\chi_{\rm soc}\,,
\label{eq:ChiExpression}
\eea

\noi where we introduced:
\begin{align}
\chi_{\rm orb}=e\upsilon_Fg_{\rm orb}\qquad{\rm and}\qquad \chi_{\rm soc}=e\upsilon g_{\rm soc}\,.
\end{align}

\noi The above are further expressed in terms of the Fermi velocity $\upsilon_F=\hbar k_F/m$ and the Fermi wavenumber $k_F=\sqrt{2m|\mu|}/\hbar$. In addition, repeated index summation ($a,b=x,y$) is implied and we made use of the Levi-Civita symbol $\varepsilon_{zab}$. In the definition of the coefficients $g_{\rm soc}$ and $g_{\rm orb}$, we also employed the compact notation $K=(\bm{k},\epsilon)$ and $\int d^3K\ph\equiv\int d\bm{k}\int_{-\infty}^{+\infty}d\epsilon$. The bare Green function {\color{black}$\hat{{\cal G}}_0(K)$ is defined in terms of the translationally invariant bare Hamiltonian $\hat{{\cal H}}_0(\hat{\bm{p}})$
through the relation:
\begin{align}\hat{{\cal G}}_0^{-1}(K)=i\epsilon-\hat{{\cal H}}_0(\bm{k}),
\end{align}

\noi where we used the Fourier transform $\hat{\bm{p}}\mapsto\hbar\bm{k}$, with $\bm{k}$ denoting the wave vector.}

\section{Topological insulator}\label{Sec:TI}

We first investigate the abovementioned interconversion effects for a single helical Dirac cone on the surface of a 3D TI. To describe this si\-tua\-tion, we take the limit $m\rightarrow\infty$ in Eq.~\eqref{eq:H0} and then find that $\chi=e\upsilon g_{\rm soc}$. Inte\-re\-stin\-gly, in this case, $\chi^{\rm TI}$ can be rewritten as:
\begin{align}
\chi^{\rm TI}=\frac{e\varepsilon_{zab}}{4\pi\hbar}\int\frac{d^3K}{(2\pi)^2}{\rm Tr}\bigg[\frac{\tau_z\hat{{\cal G}}_0(K)\sigma_z}{2i}\frac{\partial\hat{{\cal G}}_0(K)}{\partial k_a}\frac{\partial\hat{{\cal G}}_0^{-1}(K)}{\partial k_b}\bigg].
\label{eq:chiTI}
\end{align}

\noi We proceed by parametrizing the Green function as:
\bea
\hat{{\cal G}}_0(K)=\hat{\cal U}(\bm{k})\hat{G}_0(k,\epsilon)\hat{\cal U}^{\dag}(\bm{k})
\eea

\noi with $\hat{\cal U}(\bm{k})={\rm Exp}[i\vartheta(\bm{k})\sigma_z/2]$, where we defined the angle $\tan\vartheta(\bm{k})=-k_x/k_y$, the mo\-du\-lus $k=|\bm{k}|$, and the rotated frame Green function:
\bea
\hat{G}_0^{-1}(k,\epsilon)=i\epsilon-\tau_z\big(\upsilon\hbar k\sigma_x-\mu\big)-\Delta\tau_x\,.
\eea

\noi Using the above formulation, we find that $\chi^{\rm TI}$ consists of two contributions, i.e., $\chi^{\rm TI}=\chi^{\rm TI, I}+\chi^{\rm TI, II}$, where:
\begin{align}
\chi^{\rm TI,s}=-\frac{e}{4\pi\hbar}\int_0^{k_c}dk\int_{-\infty}^{+\infty}\frac{d\epsilon}{2\pi}\frac{1}{2}{\rm Tr}\left[\hat{F}^{\rm s}(k,\epsilon)\frac{\partial \hat{G}_0(k,\epsilon)}{\partial k}\right].
\end{align}

\noi The two matrix functions appearing for $s={\rm I,II}$ read as:
\bea
\hat{F}^{\rm I}(k,\epsilon)=\tau_z\quad&{\rm and}&\quad
\hat{F}^{\rm II}(k,\epsilon)={\cal D}(k,\epsilon)+\sigma_z{\cal D}(k,\epsilon)\sigma_z\,,\no
\eea

\noi where we introduced the quantity:
\begin{align}
{\cal D}(k,\epsilon)=\hat{G}_0^{-1}(k,\epsilon)\big[\tau_z,\hat{G}_0(k,\epsilon)\big]/2.\no
\end{align}

{\color{black}We observe that, in contrast to $\chi^{\rm TI, I}$, the contribution $\chi^{\rm TI, II}$ is nonzero only in the superconductive phase. In the following subsections we precisely obtain the quantity $\chi^{\rm TI}$ by means of ana\-ly\-ti\-cal methods, and explore its behavior under various conditions.}

\subsection{Quantization Effects for a Pristine Dirac Cone in the Nonsuperconductive Regime}

At zero temperature and pairing gap value{\color{black}, i.e. $\Delta=0$,} the outcome for $\chi^{\rm TI}$ takes the transparent form:
\begin{align}
\chi^{\rm TI}(\Delta=0)=
% \frac{e{\rm sgn}(\mu)}{4\pi\hbar}\ointctrclockwise_{\cal C}\frac{d\bm{k}}{2\pi}\cdot\frac{d\vartheta}{d\bm{k}}\equiv
{\rm sgn}(\mu)\frac{e}{4\pi\hbar}\ointctrclockwise_{\cal C}\frac{d\vartheta}{2\pi}\,,\label{eq:VorticityTI}
\end{align}

\noi where the closed loop ${\cal C}$ encloses the Dirac point. From the above we conclude that $\chi$ is proportional to the vorticity of the Dirac point and, as a result, it becomes quantized. For the present model $\ointctrclockwise_{\cal C}d\vartheta/2\pi=1$ and we find: 
\bea
\chi^{\rm TI}(\Delta=0)=\chi^{\rm TI,I}(\Delta=0)={\rm sgn}(\mu)\frac{e}{4\pi\hbar}\,.
\label{eq:QuantizationRelation}
\eea

Notably, the above result is generalizable to the wider class of ${\cal T}$-invariant semimetals (SMs) with an odd-parity SOC of the form $d_x(\bm{k})\sigma_y-d_y(\bm{k})\sigma_x$. In this case, $\vartheta(\bm{k})$ has to be redefined as $\tan[\vartheta(\bm{k})]=-d_y(\bm{k})/d_x(\bm{k})$. Hence, for a SM band structure with touching points, we have:
\begin{align}
\chi^{\rm SM}(\Delta=0)=\frac{e}{4\pi\hbar}\sum_s{\color{black}\omega}_s{\rm sgn}(\mu_s)
\end{align}

\noi where ${\color{black}\omega}_s$ is the vorticity of the $s$-th touching point and $\mu_s$ the respective chemical potential value which controls its occupation.

{\color{black} At this point, we also note that for a zero pai\-ring gap $\Delta$, Eq.~\eqref{eq:chiTI} can be rewritten in an alternative and more transparent form, which further highlights the to\-po\-lo\-gi\-cal origin and robustness of $\chi^{\rm TI}(\Delta=0)$. By making use of the relation $\partial\hat{{\cal G}}_0=-\hat{{\cal G}}_0\big(\partial\hat{{\cal G}}_0^{-1}\big)\hat{{\cal G}}_0$ and the property $[\tau_z,\hat{{\cal G}}_0(K)]=0$, which is a consequence of the electric charge conservation which holds for $\Delta=0$, we find that:
\begin{align}
\chi^{\rm TI}=\frac{e\varepsilon_{zab}}{4\pi\hbar}\int\frac{d^3K}{(2\pi)^2}{\rm Tr}\bigg[\frac{\tau_z\sigma_z}{2}\frac{\partial \hat{{\cal G}}_0^{-1}}{\partial\epsilon}\hat{{\cal G}}_0\frac{\partial\hat{{\cal G}}_0^{-1}}{\partial k_a}\hat{{\cal G}}_0\frac{\partial\hat{{\cal G}}_0^{-1}}{\partial k_b}\hat{{\cal G}}_0\bigg]
\label{eq:chiTItopoALT}
\end{align}

\noi where we omitted the argument $K$ from the matrix Green function $\hat{{\cal G}}_0(K)$ for compactness.

Noteworthy, the expression of the interconversion coefficient in Eq.~\eqref{eq:chiTItopoALT} has a similar structure to the coefficients controlling topological spin transport in $^3$He~\cite{Volovik1989}. Despite the fact that also in the latter case the spin current is driven by magnetization gradients, the phe\-no\-me\-na discussed in Ref.~\onlinecite{Volovik1989} are not mediated by Rashba SOC as in the cases that we study here. Nonetheless, the si\-mi\-la\-ri\-ties regarding the structure of the transport coefficients further corroborate the topological nature of the effect also in the present context. At this stage, it is also important to stress that for $\Delta=0$ and $\mu=0$, the matrix $\tau_z\sigma_z$ leads to a chiral symmetry since in this case $\tau_z\sigma_z\hat{{\cal G}}_0(K)\tau_z\sigma_z=-\hat{{\cal G}}_0^*(K)$. This property also implies the relation $\chi^{\rm TI}(\Delta=0,\mu)=-\chi^{\rm TI}(\Delta=0,-\mu)$ which ensures that $\chi^{\rm TI}(\Delta=\mu=0)=0$ as found in Eq.~\eqref{eq:QuantizationRelation}.
}

\subsection{Orbital Magnetization Picture and Realization of Parity Anomaly at Finite Density}\label{sec:OM}

In this section, we show that the result of Eq.~\eqref{eq:QuantizationRelation} is also derivable using the modern theory of orbital magnetization~\cite{OrbitalMag,Niu}, which is here denoted ${\cal M}_z(\bm{r})$. The latter is read out by expressing the energy density as:
\begin{align}
E(\bm{r})=-{\cal M}_z(\bm{r})B_z(\bm{r})
\end{align}

\noi which further implies the following alternative defining relation for the interconversion coefficient:
\begin{align}
\chi=-\left.\frac{\delta {\cal M}_z(\bm{r})}{\delta M_z(\bm{r})}\right|_{M_z(\bm{r})=0}.
\end{align}

It is more convenient to calculate $\chi$ by means of consi\-de\-ring a uniform $M_z$ which, in turn, also yields a uniform orbital magnetization. In this case, the functional derivative in the above definition simplifies to a standard partial derivative, i.e., $\delta {\cal M}_z(\bm{r})/\delta M_z(\bm{r})\mapsto\partial{\cal M}_z/\partial M_z$.

Since in the normal phase of the Rashba systems of interest the band structure consists of two bands, the expression for the corresponding uniform orbital magnetization at $T=0$ is well known~\cite{Niu}, and reads as:
\begin{align}
{\cal M}_z=\frac{e}{4\pi\hbar}\sum_{\nu=\pm1}2\mu\nu\int\frac{d\bm{k}}{2\pi}\ph\Omega(\bm{k})\Theta\big[\mu+\nu E(\bm{k})\big]\,,
\end{align}

\noi where $\Theta$ denotes the Heaviside unit step function. The above is obtained by focusing on the electron part of the respective bare Hamiltonian:
\begin{align}
\hat{{\cal H}}_{0;\tau_z=1}^{\Delta=0,M_z}(\bm{k})=\upsilon\hbar(k_x\sigma_y-k_y\sigma_x)-M_z\sigma_z-\mu\,,
\end{align}

\noi which we parametrize according to the following compact manner: $\hat{{\cal H}}_{0;\tau_z=1}^{\Delta=0,M_z}(\bm{k})\equiv\bm{d}(\bm{k})\cdot\bm{\sigma}-\mu$, where:
\begin{align}
\bm{d}(\bm{k})=\big(-\upsilon\hbar k_y,\upsilon\hbar k_x,-M_z\big)\,.
\end{align}

\noi The above vector possesses the modulus $E(\bm{k})\equiv|\bm{d}(\bm{k})|$. In addition, $\Omega(\bm{k})$ corresponds to the Berry curvature of the valence band, and is given as:
\begin{align}
\Omega(\bm{k})=\frac{1}{2}\hat{\bm{d}}(\bm{k})\cdot\left[\frac{\partial\hat{\bm{d}}(\bm{k})}{\partial k_x}\times\frac{\partial\hat{\bm{d}}(\bm{k})}{\partial k_y}\right]=-\frac{(\upsilon\hbar)^2M_z}{2E^3(\bm{k})},
\end{align}

\noi where $\hat{\bm{d}}(\bm{k})=\bm{d}(\bm{k})/E(\bm{k})$.

The desired coefficient is thus obtained via the expression $\chi^{\rm TI}(\Delta=0)=-\left.\partial{\cal M}_z/\partial M_z\right|_{M_z=0}$. In agreement with the results obtained in Refs.~\onlinecite{NiemiSemenoffRep,Sissakian}, and more recently in Ref.~\onlinecite{HuangKotetes}, we find the orbital magnetization:
\begin{align}
{\cal M}_z=-\frac{e}{4\pi\hbar}\left[\frac{\Theta\big(|M_z|-|\mu|\big)}{|M_z|}+\frac{\Theta\big(|\mu|-|M_z|\big)}{|\mu|}\right]\mu M_z\no\,.
\end{align}

From the above result, we conclude that for $|\mu|<|M_z|$ we obtain ${\cal M}_z=-{\rm sgn}(M_z)e\mu/4\pi\hbar$, which reflects the rea\-li\-za\-tion of the phenomenon of parity anomaly which is characteristic of a Dirac electron defined in two spatial dimensions with a mass $M_z$, cf Ref.~\onlinecite{Redlich}. In this regime, the system lies in its insulating phase and parity anomaly can be understood in terms of the fractional quantization of the {\color{black}anomalous} Hall conductance which is defined through the expression~\cite{Niu}:
\begin{align}
\sigma_H=-e\frac{\partial{\cal M}_z}{\partial\mu}={\rm sgn}(M_z)\frac{e^2}{4\pi\hbar}={\rm sgn}(M_z)\frac{G_0}{2}
\end{align}

\noi where $G_0=e^2/h$ is the unit of conductance and $h$ is the Planck constant. The effect is termed anomalous because $\sigma_H$ depends only on the sign of the Dirac mass $M_z$ and not its magnitude.

In this work, we are instead inte\-re\-sted in the limit $|\mu|>|M_z|$, since $M_z$ is considered to be a weak perturbation. In this regime, the theory of orbital magnetization reproduces as expected the expression in Eq.~\eqref{eq:QuantizationRelation}. Remarkably, we demonstrate that even in the metallic phase one can define a quantized quantity which plays {\color{black}an analogous role to} $\sigma_H$ in the insulating phase. Indeed, the quantized quantity here is given by:
\begin{align}
\chi=-\left.\frac{\partial{\cal M}_z}{\partial M_z}\right|_{M_z=0},
\end{align}

\noi i.e., the derivative of the orbital magnetization with respect to the Dirac mass. Hence, our central result is that massive/massless Dirac electrons exhibit a quantized response even in the metallic phase, which is now anomalous in the sense that the quantized quantity (interconversion coefficient in our context) satisfies $\chi\propto{\rm sgn}(\mu)$.

\subsection{Landau Level Picture for Uniform Magnetization Gradients}

{\color{black}In the special case of uniform magnetization gradients, the arising spin and current responses can be understood through the emergence of gapful and gapless Landau level bands.}

\subsubsection{In-plane Magnetization Gradients}

We first consider a spatial gradient for the in-plane magnetization of the form $M_y(y)={\cal B}y$. Without loss of generality, we consider that the slope of the abovementioned spatial profile is positive, i.e., ${\cal B}=\partial_yM_y>0$. In addition, we introduce an auxiliary uniform out-of-plane magnetization component $M_z$. The latter will allow us to calculate the induced out-of-plane spin density $S_z$ and, in turn, the coefficient $g_{\rm soc}^{\rm TI}=\chi^{\rm TI}/e\upsilon$. The respective Hamiltonian takes the following form:
\bea
\hat{{\cal H}}_{0;\tau_z=1}^{\Delta=0,M_z}(\hat{p}_y,y,k_x)&=&\big(\upsilon\hbar k_x-{\cal B}y\big)\sigma_y-\upsilon\hat{p}_y\sigma_x\no\\
&&-M_z\sigma_z-\mu.
\eea

\noi Following standard methods which are detailed in Appendix~\ref{app:LandauLevel}, we find the energy dispersions of the system:
\bea
E_0(k_x,M_z)&=&M_z-\mu\quad{\rm and}\\
E_{n,\pm}(k_x,M_z)&=&\pm E_n(M_z)-\mu
\eea

\noi with $n\in\mathbb{N}^+$. In the above, we introduced the energy levels for $n\geq1$:
\begin{align}
E_n(M_z)=\sqrt{(\hbar\omega_{\cal B})^2n+M_z^2}
\end{align}

\noi with the frequency $\omega_{\cal B}=\sqrt{2}\upsilon/\ell_{\cal B}$ and the lengthscale $\ell_{\cal B}=\sqrt{\upsilon\hbar/{\cal B}}$. {\color{black}The arising Landau-level band structure is depicted in Fig.~\ref{fig:Figure2}(a). As we explain in Appendix~\ref{app:LandauLevel}, the ge\-ne\-ra\-tion of the out-of-plane spin density $S_z$ accor\-ding to Eq.~\eqref{eq:SpinDensity}, solely results from the presence of the zeroth-Landau level corresponding to the quantum number with value $n=0$.}

\begin{figure*}[t!]
\begin{center}
\includegraphics[width=\textwidth]{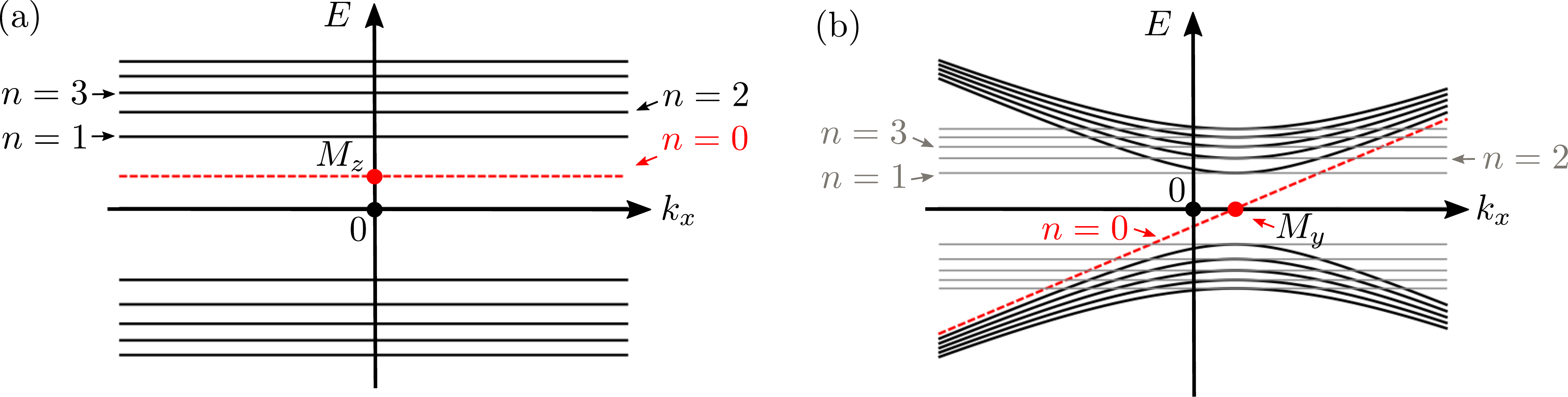}
\end{center}
\caption{{\color{black}In panel (a) [(b)] we depict the energy dispersions obtained when the nonsuperconducting surface states of the 3D TI are under the influence of a uniform magnetization gradient in the $M_y$ ($M_z$) component, which becomes inhomogeneous along the $y$ axis. The chemical potential is set to zero for convenience. The system is additionally coupled to a uniform magnetization component $M_z$ ($M_y$) which is employed to evaluate the induced uniform spin density $S_z$ ($S_y$). For a uniform gradient of the in-plane magnetization, the induced spin density $S_z$ solely results from the $n=0$ Landau level when $M_z\rightarrow0$. Similarly, in the limit $M_y\rightarrow0$ and a uniform gradient in the out-of-plane magnetization, the induced spin density $S_y$ stems only from the $n=0$ energy band which is now dispersive. For the plots we used $M_y=M_z=0.5\hbar\omega_{\cal B}$, $\hbar\omega_{\cal B}=1$, $\mu=0$, and $\upsilon\hbar k_x\in[-1,1]$.}}
\label{fig:Figure2}
\end{figure*}

\subsubsection{Out-of-plane Magnetization Gradients}

We now proceed with examining the other possible scenario, that is, to have a spatially varying out-of-plane magnetization. In the following, we consider the concrete profile $M_z(y)={\cal B}y$, where now ${\cal B}=\partial_yM_z$. Without any loss of generality, ${\cal B}$ is considered positive in the analysis below. Since for such a magnetization profile we expect the generation of a uniform spin density $S_y$, we also consider the presence of a uniform inplane magnetization $M_y$. Thus, the Hamiltonian describing this situation now becomes:
\bea
\hat{{\cal H}}_{0;\tau_z=1}^{\Delta=0,M_y}(\hat{p}_y,y,k_x)&=&\big(\upsilon\hbar k_x-M_y\big)\sigma_y-\upsilon\hat{p}_y\sigma_x\no\\&&-{\cal B}y\sigma_z-\mu\,.
\eea

Similar to the previous case, also here, the Hamiltonian is described by a spectrum of the form $E_0(k_x,M_y)$ and $E_{n,\pm}(k_x,M_y)$ with $n\in\mathbb{N}^+$. However, as we discuss in more detail in Appendix~\ref{app:LandauLevel}, the dispersions here have a different structure as one observes from the expressions shown below:
\bea
E_0(k_x,M_y)&=&\upsilon\hbar k_x-M_y-\mu\quad{\rm and}\\
E_{n,\pm}(k_x,M_y)&=&\pm E_n(k_x,M_y)-\mu,\\\no
\eea

\noi in which we introduced the energies for $n\geq1$:
\begin{align}
E_n(k_x,M_y)=\sqrt{(\hbar\omega_{\cal B})^2n+\big(\upsilon\hbar k_x-M_y\big)^2}\,.
\end{align}

Indeed, we observe that significant differences arise between the spectra of the two different types of magnetization gradients. {\color{black}See for a comparison panels (a) and (b) of Fig.~\ref{fig:Figure2}. The most notable difference is that for the in-plane case we obtained a collection of flat bands,} while in the out-of-plane situation the resulting bands are dispersive. Since the eigenenergies are even under $k_x\leftrightarrow-k_x$ for $M_y=\mu=0$, one finds that similar to the previous paragraph, also here, it is the mode associated with the $n=0$ level which contributes to the response of the system. In the present case, the $n=0$ corresponds to a chiral mode that appears at the boundary of a quantum anomalous Hall insulator described by the Hamiltonian $\propto \upsilon\hbar(k_x\sigma_y-k_y\sigma_x)-M_z\sigma_z-\mu$. This boundary is defined as the line across which $M_z$ changes sign.

{\color{black}Concluding this paragraph, we remark that for uniform gradients of the out-of-plane magnetization alternative adiabatic approaches also apply. These are discussed in Appendix~\ref{app:Adiab}.}

\subsection{Deviations away from a Pristine Dirac Cone}

We now consider deviations from the ideal Dirac cone structure of the TI surface states in the nonsuperconducting case, since it is crucial to assert the degree of robustness of the quantization of the interconversion coefficient. Specifically, we now allow for the effective mass $m$ to be finite, but yet set it to be sufficiently large so not to lead to an additional helical branch. Evenmore, we add possible warping terms, cf Ref.~\onlinecite{Mendler} {\color{black}and references therein}, which enter in the TI surface states Hamiltonian through the term $\gamma k_x(k_x^2-3k_y^2)\sigma_z$. The latter is expressed in the basis which does not include pairing. In the presence of these two terms, $\chi^{\rm TI}$ and $g_{\rm soc}^{\rm TI}$ are no longer expected to be proportional, i.e., $\chi^{\rm TI}\neq e\upsilon g_{\rm soc}^{\rm TI}$.

It is thus interesting to explore whether the quantization effects at $\Delta=0$ persist when $1/m$ and $\gamma$ are small and their effects can be examined perturbatively. We perform related calculations in Appendices~\ref{app:Dev} and~\ref{app:Dev2}, using the methods of linear response and the orbital magnetization approach, respectively.

We indeed show that inclu\-ding a quadratic kinetic ener\-gy $(\hbar\bm{k})^2/2m$ and a warping term $\gamma k_x(k_x^2-3k_y^2)\sigma_z$ leads to $\chi^{\rm TI}\neq e\upsilon g_{\rm soc}^{\rm TI}$. In more detail, a perturbative expansion in $\gamma$ and $1/m$ yields that $g_{\rm soc}^{\rm TI}(\Delta=0)$ is affected by these only at second order or higher. Specifically, up to second order we obtain the following expression:
\begin{align}
g_{\rm soc}^{\rm TI}(\Delta=0)=\frac{{\rm sgn}(\mu)}{4\pi\upsilon\hbar}\left\{1+\frac{1}{2}\left(\frac{\mu}{m\upsilon^2}\right)^2-\frac{3}{2}\left[\frac{\gamma\mu^2}{\big(\upsilon\hbar\big)^3}\right]^2\right\}.
\label{eq:gspinCorrected}
\end{align}

In stark contrast, $\chi^{\rm TI}(\Delta=0)$ becomes already mo\-di\-fied at first order in $1/m$ by an amount of $\chi^{\rm TI}(\Delta=0)\mu/m\upsilon^2$. In particular, by retaining up to second order terms in $\gamma$ and $1/m$, we find:
\begin{align}
\chi^{\rm TI}(\Delta=0)
=e\upsilon g_{{\color{black}\rm soc}}^{\rm TI}(\Delta=0)+\frac{e}{4\pi\hbar}\frac{|\mu|}{m\upsilon^2}\left(1-\frac{\mu}{m\upsilon^2}\right).
% =\frac{e}{4\pi\hbar}{\rm sgn}(\mu)\left\{1+\frac{\mu}{m\upsilon^2}-\frac{1}{2}\left(\frac{\mu}{m\upsilon^2}\right)^2-\frac{3}{2}\left[\frac{\gamma\mu^2}{\big(\upsilon\hbar\big)^3}\right]^2\right\}
\label{eq:chiCorrectedMain}
\end{align}

Notably, besides the part of $\chi^{\rm TI}(\Delta=0)$ which is proportional to the modified $g_{{\color{black}\rm soc}}^{\rm TI}(\Delta=0)$, the additional term affects $\chi^{\rm TI}(\Delta=0)$ at first order with respect to $1/m$, due to the vertex correction $\propto\bm{k}/m$. See Appendix~\ref{app:Dev} for additional related details.

In conclusion, the dichotomy emerging from the above results reveals that the interconversion coefficient $g_{\rm soc}^{\rm TI}(\Delta=0)$ is better protected than the coefficient $\chi^{\rm TI}(\Delta=0)$. Hence, measuring the arising spin density appears as a more robust route in order to experimentally identify these phenomena.

\subsection{Accounting for the Zeeman Effect}\label{sec:ZeemanEffects}

Up to now, we have fully neglected the consequences of the Zeeman effect. As we showed in Sec.~\ref{sec:SymmetryClass}, the interconversion coefficient $\chi$ becomes already modified for an arbitrarily weak value of the Land\'e factor $g$, therefore implying that it is important to examine the robustness of the quantized phenomena found here. Notably, however, $g_{\rm soc}$ is not affected by the Zeeman coupling.

In order to infer the degree of the quantization of $\chi^{\rm TI}(\Delta=0)$ for a pristine Dirac cone against the de\-via\-tions introduced by the Zeeman coupling to the external field, it is required to calculate the out-of-plane static spin susceptibility $\chi_\perp^{\rm spin}$, which is given by the expression:
\begin{align}
\chi_\perp^{\rm spin}=-\int\frac{d^3K}{(2\pi)^3}\frac{1}{2}{\rm Tr}\left[\sigma_z\hat{\cal G}_0(K)\right]^2
%
% =-\int\frac{d\bm{k}}{(2\pi)^2}\int_{-\infty}^{+\infty}\frac{d\epsilon}{2\pi}{\rm tr}\left[\sigma_z\frac{i\epsilon+\mu+\upsilon\hbar(k_x\sigma_y-k_y\sigma_x)}{(\epsilon-i\mu)^2+(\upsilon\hbar k)^2}\right]^2
=\frac{\Lambda-|\mu|}{2\pi\big(\upsilon\hbar\big)^2}.\no
\end{align}

\noi Notably, the term $\propto\Lambda$ is spurious and needs to be dropped, since it contains unphysical contributions from regions far away from the Fermi level. A similar argument was previously invoked in Ref.~\onlinecite{KopninSoninPRB} for the calculation of the superfluid stiffness of superconducting Dirac electrons. Hence, the {\color{black}properly regularized} out-of-plane spin susceptibility is of diamagnetic nature and reads as:

\begin{align}
\chi_\perp^{\rm spin}=-\frac{|\mu|}{2\pi(\upsilon\hbar)^2}.
\end{align}

As consequence, taking into account the Zeeman effect leads to the contribution $\chi_Z=-g\mu_B|\mu|/[4\pi(\upsilon\hbar)^2]$. Considering for simplicity $g=2$, and by replacing the magneton Bohr $\mu_B$ by its defining expression, we find that:
\begin{align}
\chi_Z=-\frac{e}{4\pi\hbar}\frac{|\mu|}{m_e\upsilon^2},
\end{align}

\noi where $m_e$ denotes the electron's mass. From the above, we immediately observe that the correction due to the Zeeman coupling is of the same form as the one induced by the quadratic kinetic energy at lowest order in $1/m$, but for a mass given by $m_e$. {\color{black}See Eq.~\eqref{eq:chiCorrectedMain} for a comparison.} We thus conclude that as long as the system is tuned near the Dirac point and the strength of the SOC is sufficiently strong, the correction due to the Zeeman effects is negligible. Evenmore, for $m=m_e$ the Zeeman contribution to ${\cal X}$ becomes cancelled out by the linear order term in $\mu$ appearing in $\chi^{\rm TI}$, thus rendering the spin and current responses equally protected against these types of perturbations.

\subsection{Effect of a Superconducting Gap}

The remarkable quantization encountered above for $\chi^{\rm TI}(\Delta=0)$ is always spoiled upon introducing the pai\-ring term. Besides the fact that $\chi^{\rm TI, II}$ is rendered nonzero, corrections are also introduced to $\chi^{\rm TI, I}$ of Eq.~\eqref{eq:QuantizationRelation}, which now becomes:
\begin{align}
\frac{\chi^{\rm TI,I}(\Delta)}{\chi^{\rm TI}(\Delta=0)}=\frac{1}{\sqrt{1+\lambda^2}}\equiv\frac{1}{f_\lambda}\,,
\label{eq:chiTII}
\end{align}

\noi where $f_z=\sqrt{1+z^2}$ and $\lambda=\Delta/|\mu|$. On top of the above, one now finds the following additional contribution stemming solely from the presence of a pairing gap:
\begin{align}
\frac{\chi^{\rm TI,II}(\Delta)}{\chi^{\rm TI}(\Delta=0)}=-\frac{1}{2}\frac{1}{f_\lambda}+\frac{\lambda^2}{2}\left[\frac{1}{f_\lambda}+\ln\left(\frac{\lambda}{1+f_\lambda}\right)\right].
\end{align}

\begin{figure}[t!]
\begin{center}
\includegraphics[width=0.95\columnwidth]{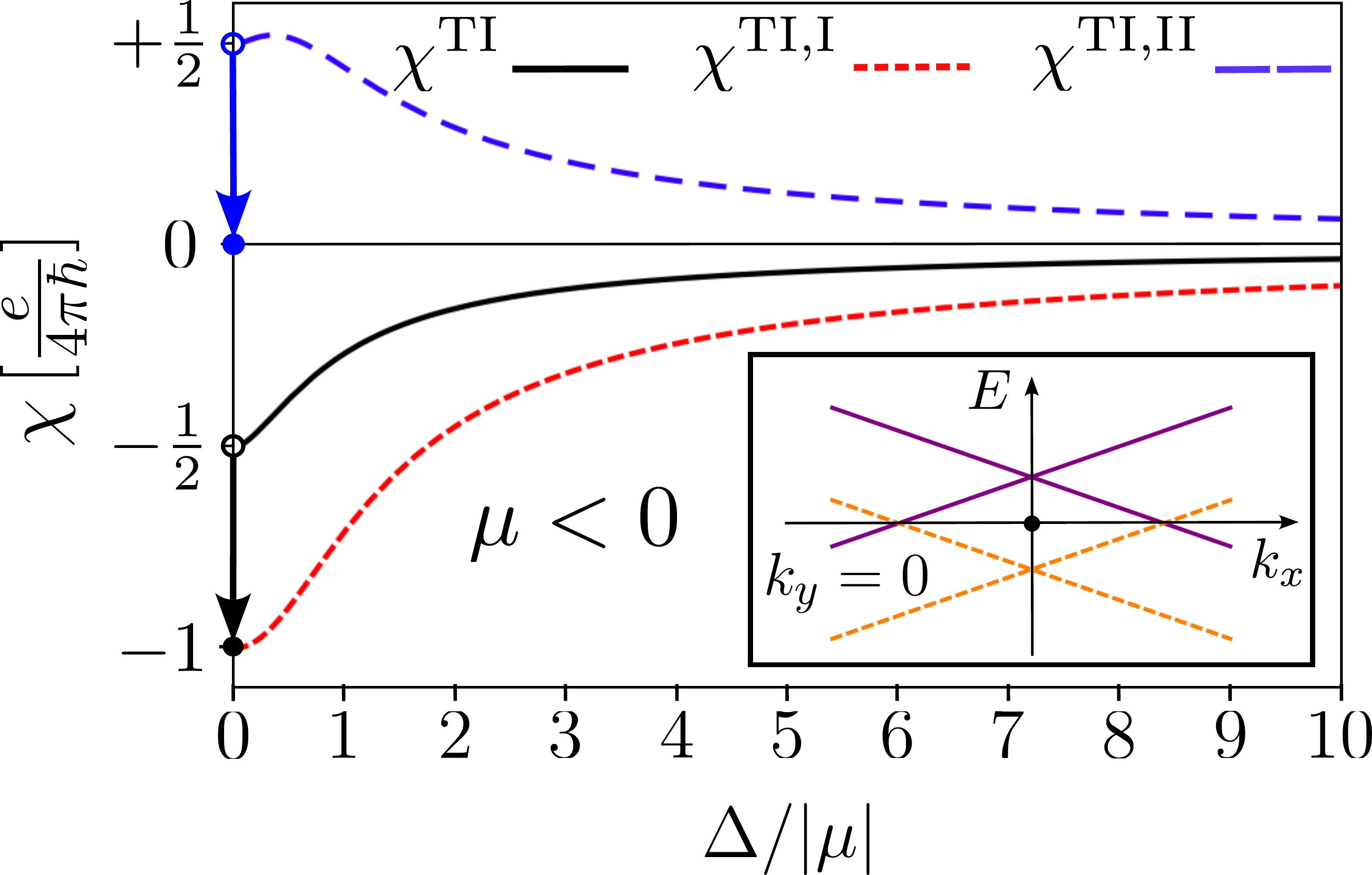}
\end{center}
\caption{Dependence of the interconversion coefficients for a 2D TI surface harboring a single pristine Dirac cone. The coefficient $\chi^{\rm TI}=e\upsilon g_{{\color{black}\rm soc}}^{\rm TI}$ consists of the parts $\chi^{\rm TI,I/II}$ which exhibit quantum anomalies. First, $\chi^{\rm TI,I}$ is independent of $\upsilon$ and persists in the limit $\upsilon\rightarrow0$. Second, $\chi^{\rm TI,II}$ remains nonzero in the limit $\Delta\rightarrow0^+$. The latter leads to the jumps observed and can be attributed to the additional Dirac cones formed away from $\bm{k}=\bm{0}$ shown in the inset.}
\label{fig:Figure3}
\end{figure}

Notably, we find that the above term does not va\-nish in the limit $\Delta\rightarrow0^+$, but instead it becomes quantized according to $\chi^{\rm TI,II}(\Delta\rightarrow0^+)=-\chi^{\rm TI}(\Delta=0)/2$. The arising discontinuous behavior of $\chi^{\rm TI}$ across $\Delta=0$ implies that superconductivity leads to nonperturbative effects, and hints towards the involvement of a quantum anomaly~\cite{Fradkin,VolovikBook,Redlich}. The latter can be attributed to the emergence of additional Dirac cones upon adding superconductivity, as sketched in the inset of Fig.~\ref{fig:Figure3}. There, we further show the precise dependence of $\chi^{\rm TI}$ on $\lambda$, and find that $\chi^{\rm TI}$ is suppressed upon increasing its strength.

\section{Rashba metal}\label{sec:Rashba}

{\color{black}We now proceed with the case of a Rashba metal (M) which is one of the most widely and routinely used systems in nanoelectronics and spintronics. The consideration of a Rashba metal is here important for one more reason. Since the nonsuperconducting band structure of a Rashba metal consists of two helical branches, it can be viewed under certain conditions as a system which effectively hosts two pristine Dirac cones. This, in fact, becomes especially important when examining its topological properties and responses. As a result of this pro\-per\-ty, we anticipate that anomalies such as the ones arising for the TI are not relevant for a Rashba metal, due to the even number of Dirac fermions, and it is therefore crucial to explore what type of new physics appears in the present context for such metallic systems.}

Based on the above discussion, our main expectation is to find $g_{\rm soc}^{\rm M}=0$ when the tem\-pe\-ra\-tu\-re and pairing gap are zero. Notably, this holds only as long as the mismatch in the occupation of the two {\color{black}helical} branches is negligible. This condition is automatically satisfied in the so-called quasiclassical limit, where the chemical potential is positive and much larger than the {\color{black}SOC energy scale $E_{\rm soc}=\upsilon\hbar k_F$ and the pairing gap $\Delta$.}

Since for a Rashba metal we do not expect quantization effects analogous to the ones encountered in Sec.~\ref{Sec:TI} for TI surface states, we mainly examine the impact of a pairing gap {\color{black}and the Zeeman effect} on the interconversion coefficient. Straightforward calculations using linear response in the quasiclassical limit, which we detail in Appendices~\ref{app:Appendix3} and~\ref{app:Appendix4}, lead to {\color{black}a set of results that we append and discuss below.

\subsection{Zero Pairing Gap}

In the nonsuperconducting phase we find that:
\begin{align}
g_{\rm soc}^{\rm M}(\Delta=0)=0\,.
\end{align}

\noi The above indeed confirms that possible anomalous contributions from the two occupied helical branches cancel each other out. On the other hand, we find the following two contributions:
\begin{align}
g_{\rm orb}^{\rm M, intra}(\Delta=0)=-g_{\rm orb}^{\rm M, inter}(\Delta=0)=\frac{1}{4\pi\hbar\upsilon_F}\,.
\end{align}

\noi Hence, the above cancel each other out, further implying that $\chi^{\rm M}(\Delta=0)=0$. Therefore, within the quasiclassical limit adopted here, and by not considering the Zeeman effect, both spin and current densities are negligible. Notably, the above result holds even in the presence of weak disorder, i.e., with a strength which leads to a level broa\-dening much smaller than the spin splitting induced by the Rashba SOC. See Appendix~\ref{app:Appendix3} for additional details.

\begin{figure}[t!]
\begin{center}
\includegraphics[width=0.95\columnwidth]{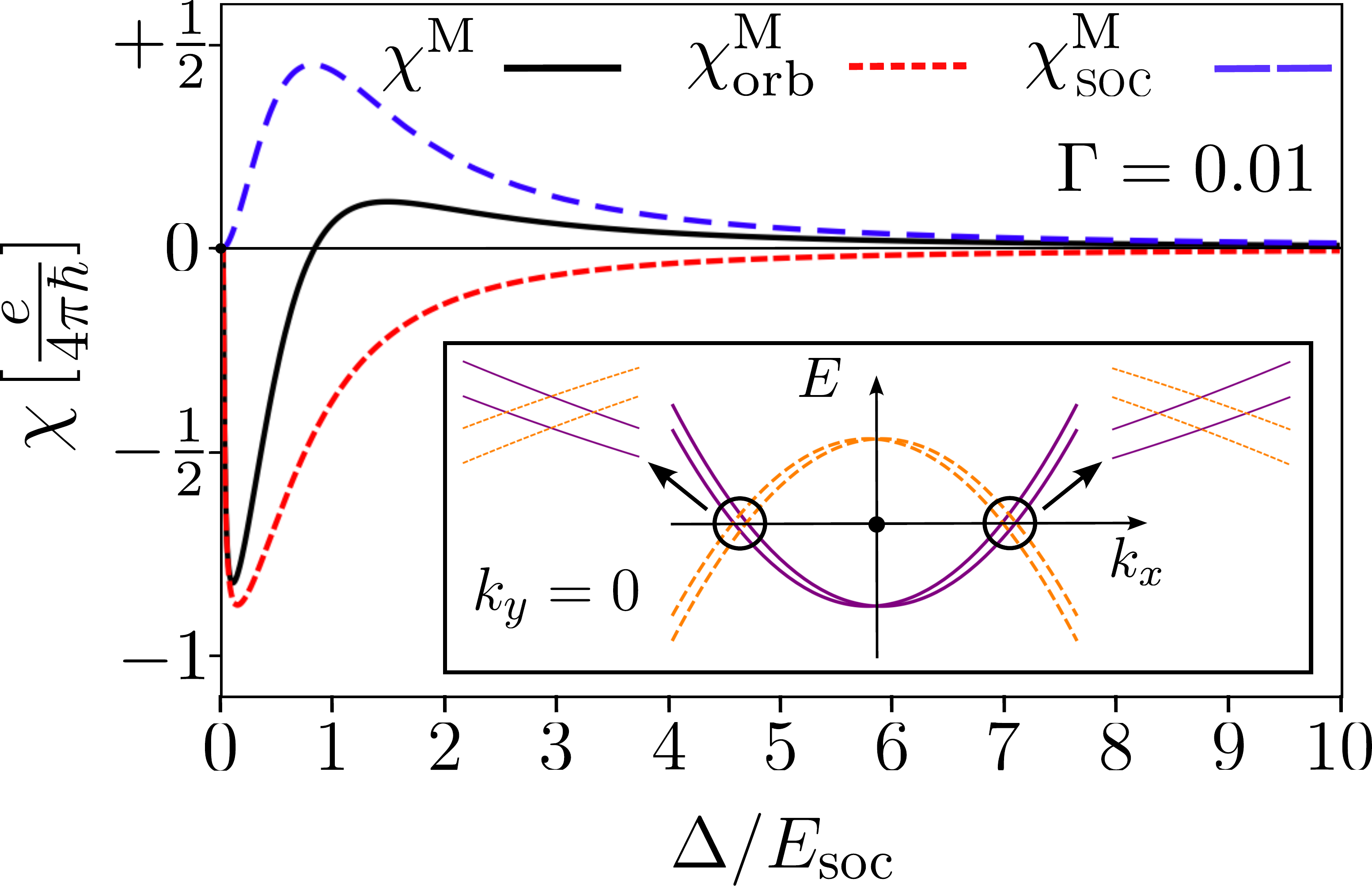}
\end{center}
\caption{Dependence of the interconversion coefficients for a Rashba metal. The coefficients here do not exhibit anomalies since the Rashba metal contains two helical branches. See inset. $\chi^{\rm M}$ is continuous in the presence of disorder, while it becomes discontinuous across $\Delta=0$ in the clean case. We used $\Gamma_\pm=\Gamma$.}
\label{fig:Figure4}
\end{figure}

\subsection{Nonzero Pairing Gap}

We now we carry out linear response in the presence of a nonzero pairing gap. Related details are once again discussed in Appendix~\ref{app:Appendix3}. Firstly, we find the following analytical result:}
\begin{align}
g_{\rm soc}^{\rm M}(\delta)=-\frac{1}{4\pi\hbar\upsilon}\left(\frac{\delta}{f_\delta}\right)^2\Bigg[1+\big(f_\delta+f_\delta^{-1}\big)\ln\left(\frac{\delta}{1+f_\delta}\right)\Bigg]
\end{align}

\noi where $\delta=\Delta/E_{\rm soc}$. The above is well-behaved in the limit $\delta\rightarrow0$, where we find $g_{\rm soc}^{\rm M}(\delta\rightarrow0)=0$. {\color{black}Therefore, $g_{\rm soc}^{\rm M}(\delta)$ is continuous upon varying $\delta$.}

The evaluation of $\chi^{\rm M}$ requires obtaining the contribution associated with $g_{\rm orb}$. Here, this coefficient retains both intra- and inter-band contributions, which read:
\bea
&&g_{\rm orb}^{\rm M, intra}(\delta)=+\frac{1}{4\pi\hbar\upsilon_F}\sum_{s=\pm}\frac{1}{2}\frac{{\color{black}\Gamma}_s}{\sqrt{{\color{black}\Gamma}_s^2+\delta^2}}\,,\label{eq:intraband}\\
&&g_{\rm orb}^{\rm M, inter}(\delta)=-\frac{1}{4\pi\hbar\upsilon_F}\left[1+\frac{\delta^2}{f_\delta}\ln\left(\frac{\delta}{1+f_\delta}\right)\right],\ph\quad
\label{eq:interband}
\eea

\noi where we phenomenologically included the broa\-de\-nings ${\color{black}\Gamma}_\pm$ as we describe in Appendix~\ref{app:Appendix3}. Note that we consider broadening effects only for the intra-band term, in order to emphasize that this contribution is sensitive even to the slightest pre\-sen\-ce of disorder and deviations away from zero tem\-pe\-ra\-tu\-re. Interestingly, an analogous behaviour dictates the intrinsic SHE~\cite{SHE,SHERMP}.

{\color{black}From Eq.~\eqref{eq:intraband}, we infer that $g_{\rm orb}^{\rm M, intra}$ is discontinuous across $\Delta=0$ for a clean system. Indeed, for $\delta\neq0$ the limit ${\color{black}\Gamma}_s\rightarrow0$ yields:
\begin{align}
g_{\rm orb,\,clean}^{\rm M, intra}(\delta\rightarrow0^+)=0\quad{\rm and}\quad
g_{\rm orb}^{\rm M,\,clean}(\delta)=-\frac{1}{4\pi\hbar\upsilon_F}\,.
\end{align}

\noi The above implies that now $g_{\rm orb}^{\rm M}(\delta\rightarrow0)=g_{\rm orb}^{\rm M, inter}$ and, thus, it becomes nonzero already in the presence of an infinitesimally weak pairing gap.

We remind the reader that in the nonsupercon\-duc\-ting regime ($\delta=0$) we obtained that $g_{\rm orb}^{\rm M, intra}$ is nonzero and, in fact, it attained a value which fully cancelled $g_{\rm orb}^{\rm M, inter}$ out. Therefore, we conclude that in the quasiclassical clean case, a nonzero pairing gap unlocks both interconversion channels, with coefficients whose dependence on $\delta$ is shown in Fig.~\ref{fig:Figure4}.

\begin{figure}[t!]
\begin{center}
\includegraphics[width=0.95\columnwidth]{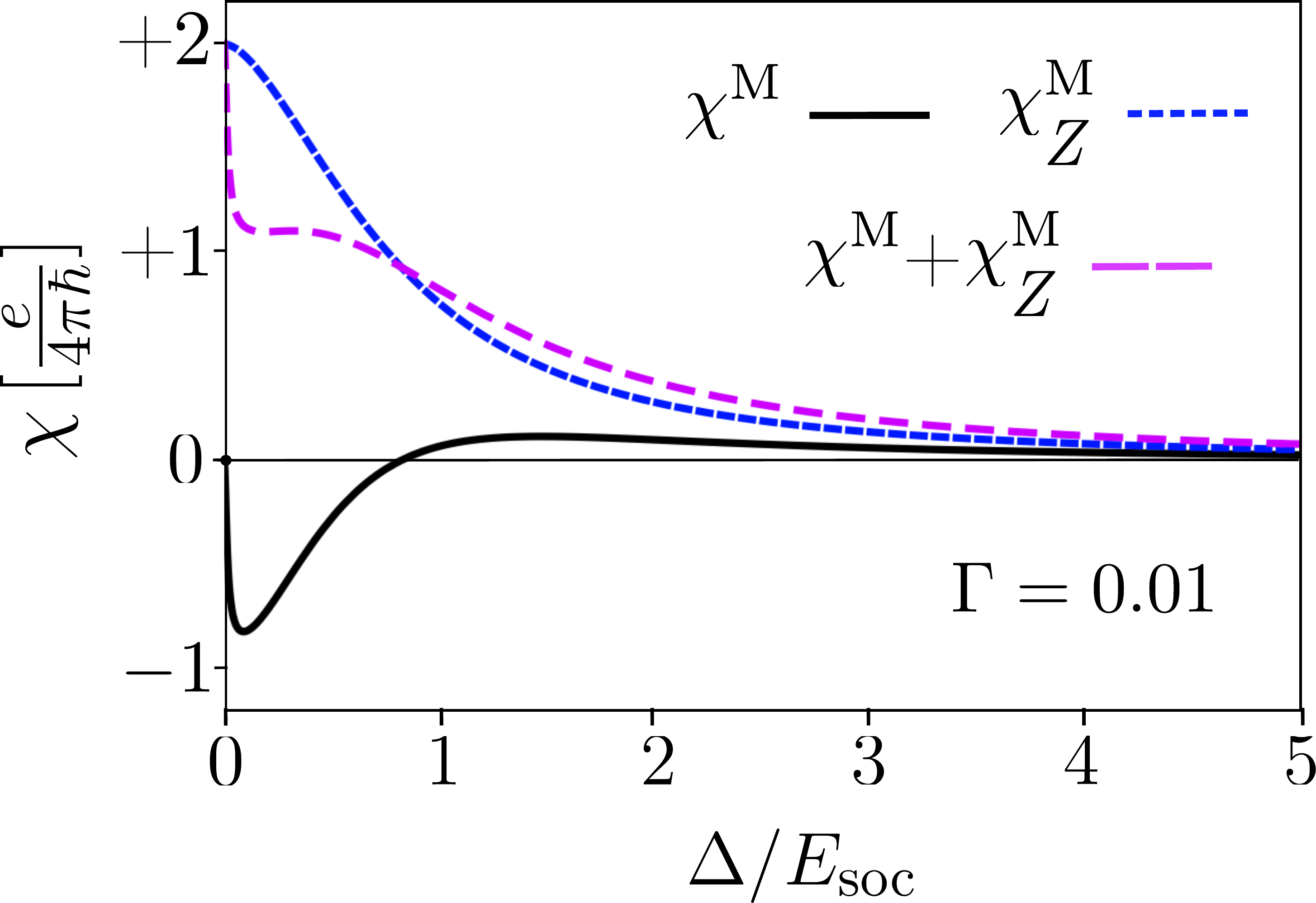}
\end{center}
\caption{{\color{black}Comparison of the interconversion coefficients $\chi$ and $\chi_Z$ for a Rashba metal, upon varying the ratio $\Delta/E_{\rm soc}$ Here, the inclusion of the Zeeman contribution $\chi_Z$ to the interconversion coefficient ${\cal X}$ is crucial, since for the former is comparable to and even larger than $\chi$. We observe that the Zeeman effect renders ${\cal X}$ positive in the entire parameter regime. For the numerical evaluation we used $\Gamma_\pm=\Gamma$.}}
\label{fig:Figure5}
\end{figure}

\subsection{Impact of the Zeeman Effect}

In order to fully infer the dependence of ${\cal X}$, our results need to be supplemented with the Zeeman contribution. For this purpose, we first obtain the out-of-plane spin susceptibility for a Rashba metal:
\begin{align}
\chi_\perp^{\rm spin}({\rm v})=2\nu_F\left[1-\frac{\ln\big({\rm v}+\sqrt{1+{\rm v}^2}\big)}{{\rm v}\sqrt{1+{\rm v}^2}}\right],\label{eq:SpinSuscMetal}
\end{align}

\noi where $\nu_F=m/2\pi\hbar^2$ defines the normal phase density of states per spin evaluated at the Fermi level. The su\-sceptibility is more conveniently parametrized in terms of the ratio ${\rm v}=1/\delta=E_{\rm soc}/\Delta$.

One finds that for ${\rm v}=0$ the susceptibility vanishes, since the spin degree of freedom becomes fully quenched for a conventional $s$-wave superconductor at zero tem\-pe\-ra\-tu\-re. Introducing instead a nonzero Rashba SOC lifts $\chi_\perp^{\rm spin}$ from zero. When ${\rm v}\rightarrow\infty$ the susceptibility becomes equal to $2\nu_F$, i.e., reaches the normal phase value under the condition $E_F\gg E_{\rm soc}$. Given the above, we now obtain the Zeeman contribution $\chi_Z$. For a system with $g=2$ and $m=m_e$, where $m_e$ defines the electron mass, we find the following expression $\chi_Z=\mu_B\chi_\perp^{\rm spin}$. The most interesting limit is when $E_{\rm soc}\gg\Delta$ since, there, also the coefficient $\chi$ is substantial. In this case, we find:
\begin{align}
\chi_Z(E_{\rm soc}\gg\Delta)=\frac{e\hbar}{2m_e}\frac{2m_e}{2\pi\hbar^2}=2\frac{e}{4\pi\hbar}\,.
\end{align}

\noi Quite interestingly, we find that the contribution of the Zeeman effect is not only non-negligible for the present system, but it is even twice as that of $\chi$. We provide a comparison of the relative strengths of $\chi$ and $\chi_Z$ in Fig.~\ref{fig:Figure5}. Therefore, accounting for the Zeeman effect is crucial for a Rashba metal.

To this end, it is important to mention that the Zeeman effect may be negligible or strongly dominant for a Rashba 2D electron gas (2DEG), where the Fermi ener\-gy $E_F$ is comparable or even smaller to $E_{\rm soc}$ and $\Delta$. Based, on the results obtained for the helical surface states of a 3D TI in Sec.~\ref{sec:ZeemanEffects}, we expect the Zeeman effect to be negligible for a low-doped Rashba 2DEG with $g\sim2$.}

\section{Experimental Relevance}\label{sec:Experimental}

{\color{black}The findings that we put forward in the preceding sections find application in a broad range of phenomena and systems. Among others, the spin and current generation stemming from magnetization gradients discussed here appears particularly prominent for topological spintro\-nics, tailoring and utilizing magnetic textures, providing alternative routes to nonreciprocal transport and superconducting diodes, and for engineering Majorana zero modes.}

\subsection{Single-Surface Detection of Topological Helical Surface States}

Firstly, the measurement of $g_{\rm soc}^{\rm TI}$ enables the detection of the helical Dirac states appearing on a given TI surface, since this interconversion coefficient is ultimately linked to the fractionally quantized {\color{black}anomalous} Hall conductance, which is considered to be their hallmark signature. Notably, the fractional Hall conductance cannot be isolated in quantum Hall measurements, since these probe pairs of surfaces~\cite{Xue,YongChen}. However, theo\-retical works~\cite{Tse,Drew} and recent experiments~\cite{Mogi} have shown that other expe\-ri\-men\-tal signatures, such as, the Faraday and Kerr effects may be equally capable of capturing the presence of these modes on a single surface.

Our analysis provides a new prominent route for une\-qui\-vo\-cal\-ly detecting helical surface states on a given surface. In particular, this becomes possible by detecting a quantized $g_{\rm soc}^{\rm TI}$ coefficient, which is expected to be an experimentally obser\-vable phenomenon when the Fermi level is tuned near the Dirac touching point of the surface band structure. Our analysis also predicts the emergence of a quantized $g_{\rm soc}^{\rm TI}$ in semi-magnetic TIs~\cite{Mogi} and thus opens an alternative detection path for parity anomaly.

\subsection{Skyrmion-Vortex Composite Pairs}

Our results also underline the need to revisit the conditions for stabilizing Bloch skyrmion-superconducting vortex excitations, since the emergence of such composite objects is governed by the value of $\chi$~\cite{Hals}. A pre\-vious work~\cite{Pershoguba} has shown that $\chi^{\rm M}\propto\big(E_{\rm soc}/E_F\big)^2$, under the assumptions $\Delta\rightarrow0$ and a non-negligible ratio $E_{\rm soc}/E_F$.

In contrast, here we take $E_F\rightarrow\infty$, which we believe is a limit that better captures the situation rea\-li\-zed in ty\-pi\-cal metals. In this regime, we show that $\chi^{\rm M}$ exhibits a non-monotonic trend with $\Delta/E_{\rm soc}$. Hence, the value of $\Delta/E_{\rm soc}$ is crucial and needs to be optimized for achie\-ving a substantial Bloch skyrmion-vortex coupling. Furthermore, we reveal that in a TI this coupling always gets suppressed upon increasing the pairing gap. This undesired property can be detrimental for enginee\-ring Majorana zero modes.

Skyrmion-vortex excitations have been recently expe\-ri\-men\-tal\-ly achieved~\cite{Panagopoulos} without, however, using a magnetoelectric mechanism~\cite{Hals}. Instead, the stray field of the skyrmions was harnessed to induce (anti)vortices~\cite{EreminSkV}. Thus, our results reveal the pre\-viou\-sly unexplored influence of superconductivity on the coupling, and promise to guide the search for new platforms which can enable Bloch skyrmion-vortex excitations~\cite{Hals}.

\subsection{Novel Superconducting Diode Effects}

Our analysis also opens up new pathways to induce nonreciprocal currents in Rashba SCs. According to previous theo\-re\-ti\-cal works~\cite{Yuan,Daido,He}, the diode effects in Refs.~\onlinecite{Ando,Baugartner,Pal} can be possibly reconciled in terms of a Rashba SOC and a pa\-ral\-lel Zeeman field, which induce a finite-momentum helical Cooper pai\-ring~\cite{Agterberg,Kaur,Hart,Mason,YuanFu}.

\begin{figure}[t!]
\begin{center}
\includegraphics[width=0.95\columnwidth]{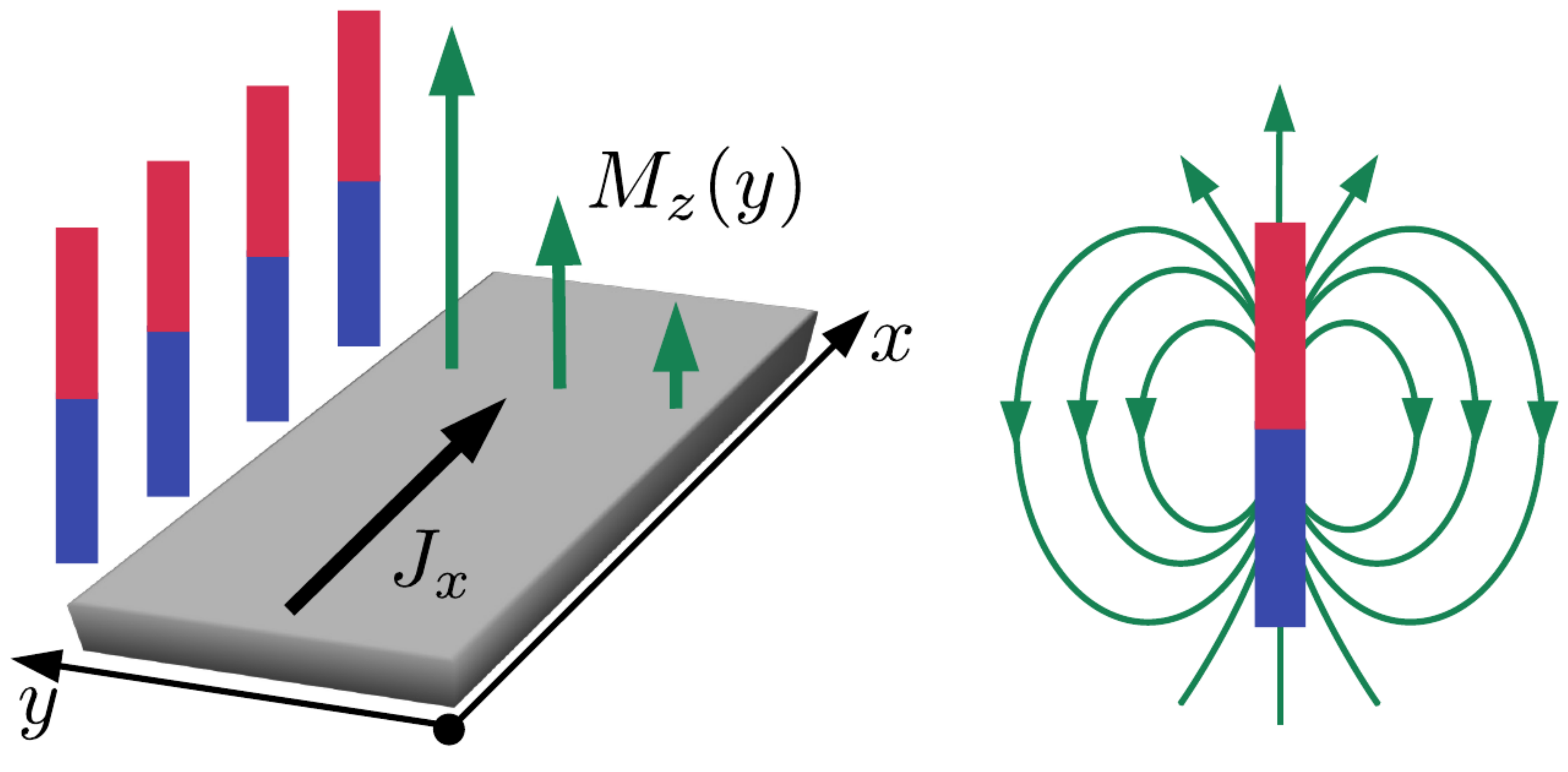}
\end{center}
\caption{Cartoon of an experimental setup for observing the superconducting diode effect in the $x$ axis, i.e., a nonreciprocal current $J_x$ due to a gradient of the out-of-plane magnetization along the $y$ axis. The magnetization profile can be induced by the fringing field of an array of nanomagnets.}
\label{fig:Figure6}
\end{figure}

Here, we predict that the diode effect can be also ge\-ne\-ra\-ted in the materials of Refs.~\onlinecite{Ando,Baugartner,Pal} by alternatively imposing a constant gra\-dient on an out-of-plane exchange field $M_z(\bm{r})$~\cite{Merce}. The latter magnetization can be engineered using, for instance, the fringing fields of a nearby array of nanomagnets~\cite{KarstenNoSOC,Mohanta}. See Fig.~\ref{fig:Figure6} for a cartoon depiction of the proposed setup.

\subsection{Majorana Zero Modes Pinned by Ferromagnetic Impurities}

Finally, our proposed effects bring forward a new route to pin superconducting vortices from an out-of-plane magnetization $M_z(\bm{r})$. In the presence of Rashba SOC, $M_z(\bm{r})$ is converted into an in-plane magnetization $\propto\bm{\nabla}M_z(\bm{r})$. This, in turn, induces an out-of-plane magnetic flux $\propto\bm{\nabla}^2M_z(\bm{r})$, which promotes vortices~\cite{HanoII}. Based on results of previous works~\cite{Gornyi,Wu,Pathak}, we infer that the ari\-sing magnetization: $\big(g_{\rm soc}\partial_xM_z(\bm{r}),g_{\rm soc}\partial_yM_z(\bm{r}),M_z(\bm{r})\big)$ is of the meron type and traps a Majorana zero mode~\cite{HanoII}.

{\color{black}
\section{Summary}\label{sec:summary}

In this work, we investigate the spin and current ge\-ne\-ra\-tion of systems with Rashba-type of spin-orbit coupling due to magnetization gradients. The latter can arise due to arrays of nanomagnets, magnetic impurities, magnetic textures and others. Our fully-analytical approach covers all the abovementioned scenarios. Specifically, we infer the interconversion coefficients which control the emergence of spin and current densities in the case of 2D Rashba metals and surfaces of 3D topological insulators.

Our study reveals that for helical surface states on a topological insulator which are dictated by a pristine Dirac cone energy spectrum, the generation of spin and current are both governed by the same interconversion coefficient. Remarkably, the latter is a topological invariant quantity, and this is reflected in the fact that it is proportional to the vorticity of the Dirac point. As we discuss, such a phenomenon is a manifestation of parity anomaly at finite density and, therefore, unveils a new class of topological effects for Dirac systems which take place in their metallic, instead of their insulating regime.

The topological nature of the interconversion coefficient further gua\-ran\-tees that this is robust against the addition of various types of weak perturbations. In particular, we first demonstrate that the inclusion of a superconducting gap always suppresses the above interconversion phenomena, which can be an obstacle for applications in the direction of topological superconductivity. Further, we addi\-tio\-nal\-ly consider the effects of the pre\-sen\-ce of a quadratic kinetic ener\-gy term and a hexagonal warping term. We find that in this case the spin and current ge\-ne\-ra\-tion are controlled by diffe\-rent coefficients, with the generation of spin being a more robust phenomenon that the induction of electrical currents. Therefore, mea\-su\-ring the spin density induced by a ferromagnetic insulator placed in pro\-xi\-mi\-ty to a single surface, promises to provide an alternative route to detect parity anomaly at finite density~\cite{Sissakian}, which appears feasible to observe in the so-called semi-magnetic topological insulators~\cite{Mogi}. Notably, pathways to detect the protected helical states on a given surface of a topological insulator are long-sought-after since observing the fractionally quantized anomalous Hall effect is practically not possible.

Besides topological insulator surfaces, our work discusses a closely related system, i.e., the 2D Rashba metal. Within a continuum description, the latter contains two Kramers degenerate points in its band structure, which in certain limits can be viewed as two Dirac touching points of opposite vorticity. Hence, the single-branch Dirac fermion anomalies encountered for a single topological insulator surface do not appear for a Rashba metal. Nonetheless, a rich va\-rie\-ty of quantization phenomena and discontinuous interconversion coefficients emerge also in the present case but exhibit different behaviors. One of the key results is that spin and current responses to magnetization gra\-dients are in general controlled by different coefficients, which exhibit discon\-ti\-nu\-ous when superconductivity is introduced to the system. Moreover, the coefficient $\chi$ controlling current generation exhibits a nonmonotonic behavior with the pairing gap, which is a property that can guide the design of hybrid systems targeted for engineering Majorana zero modes using magnetic texture defects~\cite{KlinovajaSkyrmion,Kovalev,Cren,Gornyi,Garnier,Wu,Pathak}.

The last part of our work focuses on the experimental relevance of our results. Besides the potential applications in detecting parity anomaly in alternative systems using new routes, our work is strongly relevant for si\-tua\-tions in which magnetic textures are employed as an ingredient for engineering Majorana zero modes. As a matter of fact, our work provides analytical expressions for coefficients which control the coupling between magnetic textures and superconductivity, thus enabling the improved experimental control on such systems. Even more, our study unearths new possibilities for superconducting diode effects, which reside on magnetization gradients introduced in the out-of-plane magnetization of 2D Rashba superconductors. This new possibility stems from the Rashba spin-orbit coupling instead of the usual Zeeman effect. Interestingly, such a magnetization-gradient-diode effect is in principle detectable in a number of superconductors where the diode effect has been already expe\-ri\-men\-tal\-ly but in the presence of an in-plane magnetic field~\cite{Ando,Baugartner,Pal}. Further investigations of the engineering of Majorana zero modes and the new type of diode effects proposed here will appear soon in Refs.~\onlinecite{HanoII,Merce}.
}

\section*{Acknowledgements}

We are grateful to Merc\`e Roig, Yun-Peng Huang, and Jun-Ang Wang for helpful discussions. H.~O.~M.~S. and B.~M.~A. acknowledge support from the Independent Research Fund Denmark, grant number 8021-00047B.

\appendix

\begin{widetext}

\section{Microscopic Formalism for Deriving the Interconversion Coefficients}\label{app:Appendix1}

We start from Eq.~\eqref{eq:Hamiltonian} of the main text and carry out a perturbative expansion of the energy per area of the system in terms of $M_{x,y,z}(\bm{r})$ and $A_{x,y}(\bm{r})$. In most cases, we consider the zero temperature matrix Green function framework~\cite{Fradkin,VolovikBook}, which can be obtained from the Matsubara formalism~\cite{Bruus} after taking the limit $T\rightarrow0$. $T$ is the temperature in energy units. The bare Matsubara Green function is given by $\hat{\cal G}_0^{-1}(\bm{k},ik_n)=ik_n-\hat{{\cal H}}_0(\bm{k})$, where $k_n$ correspond to fermionic Matsubara frequencies~\cite{Bruus}. In the limit $T\rightarrow0$, the Matsubara summation $T\sum_{ik_n}$ is replaced by the integral $\int_{-\infty}^{+\infty}d\epsilon /2\pi$, where $\epsilon\in(-\infty,+\infty)$.

Starting from the Matsubara formalism, we first obtain the free energy per area ${\cal F}$ up to second order in the single-particle Hamiltonian operator $\hat{V}$ which contains the perturbation. Textbook result yields that at second order the free energy acquires the contribution ${\cal F}^{(2)}=\frac{1}{2}{\rm tr}\big(\hat{\cal G}_0\hat{V}\big)^2$, where $\hat{\cal G}_0$ corresponds to the operator form of the bare Matsubara Green function, and ${\rm tr}$ denotes trace over all possible degree of freedom~\cite{Fradkin}.

At this point, we obtain the form of the perturbation term $\hat{V}$. For this purpose, we introduce the plane wave basis for the spinor in Eq.~\eqref{eq:Hamiltonian} of the main text, i.e., $\bm{\Psi}(\bm{r})=\frac{1}{2\pi}\int d\bm{k}\ph e^{i\bm{k}\cdot\bm{r}}\bm{\Psi}(\bm{k})$. We also define the Fourier transform of the vector potential and magnetization according to the general definition $a(\bm{r})=\frac{1}{(2\pi)^2}\int d\bm{q}\ph e^{i\bm{q}\cdot\bm{r}}a(\bm{q})$. The above procedure allows us to rewrite Eq.~\eqref{eq:Hamiltonian} of the main text in momentum space:
\bea
H=\frac{1}{2}\int\frac{d\bm{q}}{(2\pi)^2}\bm{\Psi}^\dag(\bm{k+q})\Big[\hat{{\color{black}{\cal H}}}_0(\bm{k})(2\pi)^2\delta(\bm{q})+\big<\bm{k}+\bm{q}\big|\hat{V}\big|\bm{k}\big>\Big]\bm{\Psi}(\bm{k})\no
\eea

\noi where we introduced the matrix elements $\big<\bm{k}+\bm{q}\big|\hat{V}\big|\bm{k}\big>$ of the perturbation term. After dropping the diamagnetic coupling to the vector potential, and neglecting at this stage possible Landau-level quantization effects, we have:
\bea
\big<\bm{k}+\bm{q}\big|\hat{V}\big|\bm{k}\big>=\left[\frac{e\hbar}{m}\left(\bm{k}+\frac{\bm{q}}{2}\right)+e\upsilon(\sigma_y,-\sigma_x)\right]\cdot\bm{A}(\bm{q})-\bm{M}(\bm{q})\cdot\bm{\sigma}\,.\no
\eea

\noi With the help of the above, we now obtain the quadratic contribution of the above perturbation to the free energy per area, which reads:
\bea
{\cal F}^{(2)}=\frac{1}{2}\int\frac{d\bm{q}}{(2\pi)^2}\int\frac{d\bm{k}}{(2\pi)^2}T\sum_{ik_n}\frac{1}{2}{\rm Tr}\left[\big<\bm{k}\big|\hat{V}\big|\bm{k}+\bm{q}\big>\hat{\cal G}_0(\bm{k}+\bm{q},ik_n)\big<\bm{k}+\bm{q}\big|\hat{V}\big|\bm{k}\big>\hat{\cal G}_0(\bm{k},ik_n)\right]\,.
\eea

\noi Note that the factor of $\nicefrac{1}{2}$ appearing in front of the trace operation ${\rm Tr}$ is introduced to avoid double counting the electronic degrees of freedom.

The desired coefficients $\chi$ and $g_{\rm orb,soc}$ are read out from the above expression. Specifically, $\chi$ is obtained from the following contribution to the free energy:
\bea
{\cal F}_{M_z;\bm{A}}^{(2)}&=&-\int\frac{d\bm{q}}{(2\pi)^2}M_z(-\bm{q})\int\frac{d\bm{k}}{(2\pi)^2}T\sum_{ik_n}\frac{1}{2}{\rm Tr}\left\{\sigma_z\hat{\cal G}_0(\bm{k}+\bm{q},ik_n)\left[\frac{e\hbar}{m}\left(\bm{k}+\frac{\bm{q}}{2}\right)+e\upsilon(\sigma_y,-\sigma_x)\right]\hat{\cal G}_0(\bm{k},ik_n)\right\}\cdot\bm{A}(\bm{q})\no\\
&\equiv&-\int\frac{d\bm{q}}{(2\pi)^2}M_z(-\bm{q})\Pi_{M_zA_a}(\bm{q})A_a(\bm{q})\,,\no
\eea

\noi where repeated index summation is above implied with $a=x,y$ and we introduced the polarization tensor:
\bea
\Pi_{M_zA_a}(\bm{q})=\int\frac{d\bm{k}}{(2\pi)^2}T\sum_{ik_n}\frac{1}{2}{\rm Tr}\left\{\sigma_z\hat{\cal G}_0(\bm{k}+\bm{q},ik_n)\left[\frac{e\hbar}{m}\left(k_a+\frac{q_a}{2}\right)+e\upsilon\varepsilon_{zab}\sigma_b\right]\hat{\cal G}_0(\bm{k},ik_n)\right\}.\no
\eea

\noi Next, we express ${\cal F}_{M_z;\bm{A}}^{(2)}$ according to:
\bea
{\cal F}_{M_z;\bm{A}}^{(2)}=\chi\int d\bm{r}\ph M_z(\bm{r})B_z(\bm{r})=\chi\int d\bm{r}\ph M_z(\bm{r})\big[\partial_xA_y(\bm{r})-\partial_yA_x(\bm{r})\big]\equiv
\chi\int\frac{d\bm{q}}{(2\pi)^2}\ph M_z(-\bm{q})\big[iq_xA_y(\bm{q})-iq_yA_x(\bm{q})\big]\no
\eea

\noi and we find that:
\bea
\chi=\frac{1}{2i}\left[\frac{\partial\Pi_{M_zA_x}(\bm{q})}{\partial q_y}-\frac{\partial\Pi_{M_zA_y}(\bm{q})}{\partial q_x}\right]_{\bm{q}=\bm{0}}\,.\label{eq:chi}
\eea

\noi The above expression directly leads to Eqs.~\eqref{eq:gspin}-\eqref{eq:ChiExpression} of the main text. A couple of related comments are in place. We first note that the component $\propto\bm{q}/2$ of the vertex $e\hbar(\bm{k}+\bm{q}/2)/m$ does not contribute to $\chi$. Second, to obtain the expressions in the main text, we take the zero temperature limit and replace the Matsubara frequencies $ik_n$ by $i\epsilon$, and the respective summations by suitable integrations.

Following a similar procedure one obtains the coefficient $g_{{\color{black}\rm soc}}$. To transparently demonstrate this, we focus on the following contribution appearing in ${\cal F}^{(2)}$:
\bea
{\cal F}_{M_z;M_{x,y}}^{(2)}&=&\int\frac{d\bm{q}}{(2\pi)^2}M_z(-\bm{q})\int\frac{d\bm{k}}{(2\pi)^2}T\sum_{ik_n}\frac{1}{2}{\rm Tr}\left\{\sigma_z\hat{\cal G}_0(\bm{k}+\bm{q},ik_n)\left[M_x(\bm{q})\sigma_x+M_y(\bm{q})\sigma_y\right]\hat{\cal G}_0(\bm{k},ik_n)\right\}\no\\
&\equiv&-\int\frac{d\bm{q}}{(2\pi)^2}M_z(-\bm{q})\chi_{M_zM_a}^{\rm spin}(\bm{q})M_a(\bm{q})\,,\no
\eea

\noi where we introduced the spin susceptibility:
\bea
\chi_{M_zM_a}^{\rm spin}(\bm{q})=-\int\frac{d\bm{k}}{(2\pi)^2}T\sum_{ik_n}\frac{1}{2}{\rm Tr}\left[\sigma_z\hat{\cal G}_0(\bm{k}+\bm{q},ik_n)\sigma_a\hat{\cal G}_0(\bm{k},ik_n)\right]\,.\no
\eea

\noi Next, we express ${\cal F}_{M_z;M_{x,y}}^{(2)}$ according to:
\bea
{\cal F}_{M_z;M_{x,y}}^{(2)}=g_{{\color{black}\rm soc}}\int d\bm{r}\ph M_z(\bm{r})\bm{\nabla}\cdot\bm{M}(\bm{r})\equiv
g_{{\color{black}\rm soc}}\int\frac{d\bm{q}}{(2\pi)^2}\ph M_z(-\bm{q})\big[iq_xM_x(\bm{q})+iq_yM_y(\bm{q})\big]\no
\eea

\noi and obtain the expression for $g_{{\color{black}\rm soc}}$:
\bea
g_{{\color{black}\rm soc}}=-\frac{1}{2i}\left[\frac{\partial\chi_{M_zM_x}^{\rm spin}(\bm{q})}{\partial q_x}+\frac{\partial\chi_{M_zM_y}^{\rm spin}(\bm{q})}{\partial q_y}\right]_{\bm{q}=\bm{0}}\,.\label{eq:gspinApp}
\eea

\noi Following the same steps as for $\chi$, the above expression provides Eq.~\eqref{eq:gspin} of the main text.

\section{Evaluation of the Interconversion Coefficients: Topological Insulator}\label{app:Appendix2}

The results presented in the main text are tedious but yet straightforward. Hence, below we provide only a few necessary remarks and clarifications concerning the derivations of the main text, while we examine further aspects regarding the stability of the quantization of the interconversion coefficients in the nonsuperconducting phase.

\subsection{Case of a Pristine Dirac Cone - Linear Response}

Regarding the TI surface states, we first assume that these are dictated by a pristine conical Dirac spectrum and neglect any possible warping effects, see for instance Ref.~\onlinecite{Mendler}. To describe such a situation, we consider that $m\rightarrow\infty$. Hence, in the present case the perturbation term becomes:
\bea
\big<\bm{k}+\bm{q}\big|\hat{V}^{\rm TI}\big|\bm{k}\big>=e\upsilon(\sigma_y,-\sigma_x)\cdot\bm{A}(\bm{q})-\bm{M}(\bm{q})\cdot\bm{\sigma}=-\big(M_x(\bm{q})+e\upsilon A_y(\bm{q}),M_y(\bm{q})-e\upsilon A_x(\bm{q}),M_z(\bm{q})\big)\cdot\bm{\sigma}\,.\no
\eea

\noi The above illustrates that the in-plane magnetization components play a role analogous to the vector potential, and results in the expression $\chi=e\upsilon g_{\rm soc}$. In the main text, we find that $\chi^{\rm TI}(\Delta=0)=\frac{e}{4\pi\hbar}{\rm sgn}(\mu)$, i.e., it is quantized. In contrast, as shown in the manuscript, the addition of a nonzero pairing gap spoils this quantization.

\subsection{Case of a Pristine Dirac Cone - Alternative Derivation using a Landau Level Approach}\label{app:LandauLevel}

In this section, we provide an alternative understanding of the quantized interconversion phenomena by residing to a Landau-level picture. For this purpose, we restrict to the case of a pristine Dirac cone in the absence of superconductivity. The cone is under the influence of a magnetization which features a constant gradient in space.

\subsubsection{Uniform Gradient for the In-Plane Magnetization}

We first consider a spatial gradient of the inplane magnetization $M_y(y)={\cal B}y$. Without loss of generality we consider that the slope of the abovementioned spatial profile is positive, i.e., ${\cal B}=\partial_yM_y>0$. In addition, we introduce an auxiliary uniform out-of-plane magnetization component $M_z$. The latter will allow us to calculate the induced out-of-plane spin density $S_z$. The respective electron part of the Hamiltonian for zero chemical potential becomes:
\begin{align}
\hat{{\cal H}}_{0;\tau_z=1}^{\Delta=\mu=0,M_z}(\hat{p}_y,y,k_x)=\big(\upsilon\hbar k_x-{\cal B}y\big)\sigma_y-\upsilon\hat{p}_y\sigma_x-M_z\sigma_z=-\big[\upsilon\hat{p}_y\sigma_x+{\cal B}\big(y-\upsilon\hbar k_x/{\cal B}\big)\sigma_y+M_z\sigma_z\big]\,.\no
\end{align}

\noi At this stage it is convenient to define the lengthscale $\ell_{\cal B}=\sqrt{\upsilon\hbar/{\cal B}}$ and the frequency $\omega_{\cal B}=\sqrt{2}\upsilon/\ell_{\cal B}$, given the choice $\upsilon>0$. With the new variables, the Hamiltonian becomes:
\begin{align}
\hat{{\cal H}}_{0;\tau_z=1}^{\Delta=\mu=0,M_z}(\hat{p}_y,y,k_x)=-\hbar\omega_{\cal B}\left[\frac{\ell_{\cal B}}{\sqrt{2}}\frac{\hat{p}_y}{\hbar}\sigma_x+\frac{1}{\sqrt{2}\ell_{\cal B}}\big(y-k_x\ell_{\cal B}^2\big)\sigma_y\right]-M_z\sigma_z\,.\no
\end{align}

\noi Following closely Ref.~\onlinecite{CastroNeto}, we now introduce the ladder operators for a given $k_x$:
\begin{align}
\hat{a}(k_x)=\frac{\ell_{\cal B}}{\sqrt{2}}\frac{\hat{p}_y}{\hbar}-i\frac{y-k_x\ell_{\cal B}^2}{\sqrt{2}\ell_{\cal B}}\quad{\rm and}\quad
\hat{a}^\dag(k_x)=\frac{\ell_{\cal B}}{\sqrt{2}}\frac{\hat{p}_y}{\hbar}+i\frac{y-k_x\ell_{\cal B}^2}{\sqrt{2}\ell_{\cal B}}
\end{align}

\noi which satisfy $[\hat{a}(k_x),\hat{a}^\dag(k_x)]=1$, and allows us to re-express the Hamiltonian according to the following form: $\hat{{\cal H}}_{0;\tau_z=1}^{\Delta=\mu=0,M_z}(\hat{p}_y,y,k_x)=-\hbar\omega_{\cal B}\big[\hat{a}^\dag(k_x)\sigma_-+\hat{a}(k_x)\sigma_+\big]-M_z\sigma_z$, where $\sigma_\pm=(\sigma_x\pm i\sigma_y)/2$. This Hamiltonian is more conveniently diagonalized by identifying the eigenstates and eigenvectors of the Hamiltonian for $M_z=0$, that reads as $\hat{{\cal H}}_{0;\tau_z=1}^{\Delta=\mu=M_z=0}(\hat{p}_y,y,k_x)=-\hbar\omega_{\cal B}\big[\hat{a}^\dag(k_x)\sigma_-+\hat{a}(k_x)\sigma_+\big]$. This Hamiltonian satisfies $\big[\hat{{\cal H}}_{0;\tau_z=1}^{\Delta=\mu=M_z=0}(\hat{p}_y,y,k_x)/\hbar\omega_{\cal B}\big]^2=\hat{a}^\dag(k_x)\hat{a}(k_x)(1-\sigma_z)/2+\hat{a}(k_x)\hat{a}^\dag(k_x)(1+\sigma_z)/2=\hat{a}^\dag(k_x)\hat{a}(k_x)+(1+\sigma_z)/2$. The latter gives rise to a zero-energy Landau level for $\sigma_z=-1$ and eigenvector:
\bea
\big|u_{0,-1}(k_x)\big>=\big|\phi_0(k_x)\big>\left(\begin{array}{c}0\\1\end{array}\right)\quad{\rm with}\quad \varepsilon_{0,-1}=0\,,\no
\eea

\noi where we introduced the eigenstates of the displaced quantum harmonic oscillator $\big|\phi_n(k_x)\big>$ (with displacement $k_x\ell_{\cal B}^2$) which satisfy the defining relation $\hat{a}^\dag(k_x)\hat{a}(k_x)\big|\phi_n(k_x)\big>=n\big|\phi_n(k_x)\big>$. Having obtained the expression for the zero-energy Landau level allows us to determine the remaining spectrum of $\hat{{\cal H}}_{0;\tau_z=1}^{\Delta=\mu=M_z=0}(\hat{p}_y,y,k_x)$, which is given in term of the following two families of non-zero-energy Landau levels:
\bea
\big|u_{n,\sigma}(k_x)\big>=\frac{1}{\sqrt{2}}\left(\begin{array}{c}\big|\phi_{n-1}(k_x)\big>\\\\\sigma\big|\phi_n(k_x)\big>\end{array}\right)\quad{\rm with}\quad \varepsilon_{n,\sigma}(k_x)=\sigma\varepsilon_n(k_x)=\sigma\hbar\omega_{\cal B}\sqrt{n}\quad{\rm for}\quad
n\geq1\,.\no
\eea

\noi Each Landau level sees a degeneracy per area which is given by $1/2\pi\ell_{\cal B}^2$. Note that due to the presence of chiral symmetry, i.e., where $\{\hat{{\cal H}}_{0;\tau_z=1}^{\Delta=\mu=M_z=0}(\hat{p}_y,y,k_x),\sigma_z\}=\hat{0}$, the eigenstates of each pair of nonzero energy Landau levels are related accor\-ding to: $\big|u_{n\geq1,\pm}(k_x)\big>=\sigma_z\big|u_{n\geq1,\mp}(k_x)\big>$. These observations allow us to immediately infer the energy spectrum when $M_z\neq0$. Specifically, we have the eigenenergies:
\begin{align}
E_0(k_x,M_z)=M_z-\mu,\qquad{\rm and}\qquad
E_{n,\pm}(k_x,M_z)=\pm E_n(M_z)-\mu\quad {\rm with} \quad E_n(M_z)=\sqrt{(\hbar\omega_{\cal B})^2n+M_z^2}\,.
\end{align}

\noi To obtain $S_z$, we need to infer the energy per area in the additional presence of a chemical potential. We have:
\begin{align}
E=\frac{1}{2\pi\ell_{\cal B}^2}
\left\{\big(M_z-\mu\big)\Theta(\mu-M_z)+\sum_{n=1}^\infty\sum_{\sigma=\pm}\big[\sigma E_n(M_z)-\mu\big]\Theta\big[\mu-\sigma E_n(M_z)\big]\right\}.
\end{align}

It is now straightforward to obtain the spin density $S_z=-\partial E/\partial M_z$. We are particularly interested in the out-of-plane spin density for $M_z=0$ which allows us to infer $g_{\rm soc}$. In this limit, it is only the zero-energy Landau level that contributes to $S_z$ since the energy contribution from the nonzero Landau levels is even under $M_z\leftrightarrow-M_z$. After replacing $\ell_{\cal B}$ by its defining relation and setting $\partial_yM_y\equiv\bm{\nabla}\cdot\bm{M}$, we find that $g_{\rm soc}^{\rm TI}$ takes the form: $g_{\rm soc}^{\rm TI}(\Delta=0)=2\Theta(\mu)/4\pi\upsilon\hbar$. At first sight, this results appears to contradict our previous conclusions since, instead of $2\Theta(\mu)$, one would instead expect the function ${\rm sgn}(\mu)$. However, there is no reason for such a distinction between $\mu>0$ and $\mu<0$ to be physical. Indeed, as it has been also pointed out in Ref.~\onlinecite{Levitov} in the frame of a different but related calculation, such a discrepancy is because a different regularization takes place in the present procedure compared to the Green function approach. Therefore, in order to properly regularize the above process, we simply antisymmetrize our result with respect to $\mu\leftrightarrow-\mu$, and find that $2\Theta(\mu)\mapsto\Theta(\mu)-\Theta(-\mu)\equiv{\rm sgn}(\mu)$. After this regularization, we recover our previously obtained result, which can now be attributed to the contribution of the zero energy Landau level.

\subsubsection{Uniform Gradient for the Out-Of-Plane Magnetization}

We now proceed with examining the other possible scenario, that is, to have a spatially varying out-of-plane magnetization. In the following, we consider the concrete profile $M_z(y)={\cal B}y$, where now ${\cal B}=\partial_yM_z$. Without any loss of generality, ${\cal B}$ is considered positive in the analysis below. Since for such a magnetization profile we expect the generation of a uniform spin density $S_y$, we also consider the presence of a uniform inplane magnetization $M_y$. Thus, the Hamiltonian describing this situation now becomes:
\begin{align}
\hat{{\cal H}}_{0;\tau_z=1}^{\Delta=\mu=0,M_y}(\hat{p}_y,y,k_x)=\big(\upsilon\hbar k_x-M_y\big)\sigma_y-\upsilon\hat{p}_y\sigma_x-{\cal B}y\sigma_z\equiv\frac{\sigma_y-\sigma_z}{\sqrt{2}}\left[\upsilon\hat{p}_y\sigma_x+{\cal B}y\sigma_y+\big(M_y-\upsilon\hbar k_x\big)\sigma_z\right]\frac{\sigma_y-\sigma_z}{\sqrt{2}}\,.\no
\end{align}

\noi By bringing the Hamiltonian in the above form, we can immediately obtain the eigenenergies of this Hamiltonian using the results of the previous paragraphs. In principle, one can obtain the resulting $S_y$ by evaluating the energy per area as we did in the previous section. However, the emergence of ${\cal B}$ in the final result is not as transparent as it was in the previous case, where it entered through the Landau level degeneracy. For this reason, we offer another route to transparently calculate $S_y$ in the presence of the constant magnetization gradient ${\cal B}=\partial_xM_z$. See Sec.~\ref{app:Adiab}.

\subsection{Adiabatic framework for Uniform Out-of-Plane Magnetization Gradients}\label{app:Adiab}

In this paragraph we discuss the quantization of the spin and current densities arising from a uniform out-of-plane magnetization gradient in terms of an alternative approach which relies on treating the inhomogeneous magnetization in a spatially adiabatic fashion. As we explained in Appendix~\ref{app:LandauLevel} and further show here, the present method allows studying the phenomena in que\-stion more transparently, while it also allows us to describe the inhomogeneous problem in a new frame, in which the Hamiltonian is translationally invariant.

For the purpose of exposing the here-proposed adia\-ba\-tic method, we re-express the Hamiltonian in the folowing manner:
\bea
\hat{{\cal H}}_{0;\tau_z=1}^{\Delta=0,M_y}(\hat{p}_y,y,k_x)&=&
\sqrt{\big(\upsilon\hbar k_x-M_y\big)^2+\big({\cal B}y\big)^2}\sigma_ye^{-i\Phi(y)\sigma_x}-\upsilon\hat{p}_y\sigma_x-\mu\no\\
&\equiv&e^{i\Phi(y)\sigma_x/2}\left\{\sqrt{\big(\upsilon\hbar k_x-M_y\big)^2+\big({\cal B}y\big)^2}\sigma_y-\upsilon\big[\hat{p}_y+\hbar\partial_y\Phi(y)\sigma_x/2\big]\sigma_x-\mu\right\}e^{-i\Phi(y)\sigma_x/2}\quad
\label{eq:Adiab}
\eea

\noi where we introduced the angle $\Phi(y)$ through the defining relation $\tan[\Phi(y)]={\cal B}y/(\upsilon\hbar k_x-M_y)$. Since $M_y$ is kept intact, we can evaluate $S_y$ using the Hamiltonian in the rotated frame appearing inside the brackets of Eq.~\eqref{eq:Adiab}:
\begin{align}
\underline{\hat{{\cal H}}}_{0;\tau_z=1}^{\Delta=0,M_y}(\hat{p}_y,y,k_x)=\sqrt{\big(\upsilon\hbar k_x-M_y\big)^2+\big({\cal B}y\big)^2}\sigma_y-\upsilon\hat{p}_y\sigma_x-\frac{\upsilon\hbar\partial_y\Phi(y)}{2}-\mu.
\end{align}

\noi For weak strengths of ${\cal B}$, i.e., which satisfy $|\mu|\gg{\cal B}L_y$, where $L_y$ denotes the sample's length in the $y$ direction, we can consider the approximation $|\upsilon\hbar k_x-M_y|\gg |y|{\cal B}$, which renders the above new-frame Hamiltonian translationally invariant. Therefore, within this adiabatic type of approach, we obtain the Hamiltonian:
\begin{align}
\underline{\hat{{\cal H}}}_{0;\tau_z=1}^{\Delta=0,M_y}(\bm{k})=|\upsilon\hbar k_x-M_y|\sigma_y-\upsilon\hbar k_y\sigma_x-\mu-\upsilon\hbar{\cal B}/\big[2(\upsilon\hbar k_x-M_y)\big]
\end{align}

\noi which reveals that the effect of the out-of-plane magnetization gradient is to effectively act as an additional $k_x$ dependent che\-mi\-cal potential which breaks inversion symmetry.

At this stage, it is important to remark that the term $\upsilon\hbar k_x-M_y$ is nonzero only as long as $\mu$ is nonzero. For ${\cal B}=0$, only one of the two he\-li\-ci\-ty branches with energies $\pm\sqrt{(\upsilon\hbar k_x-M_y)^2+(\upsilon\hbar k_y)^2}$ crosses the chemical potential. As we show below, $S_y$ originates from the Fermi level response of that single helicity branch.

To demonstrate this, we assume with no loss of ge\-ne\-ra\-li\-ty that $\mu=|\mu|>0$. Due to the anisotropic manner in which ${\cal B}$ influences the band structure, we proceed by viewing the Hamiltonian as a collection of multiple 1D systems in the $x$ direction, for which, $k_y$ plays the role of a mere parameter. In this sense, the Fermi points of the upper helicity branch which crosses the Fermi level are given by $\upsilon\hbar k_x-M_y=\pm\sqrt{\mu^2-(\upsilon\hbar k_y)^2}$, under the condition $|\upsilon\hbar k_y|\leq|\mu|$. In the following, the two Fermi points are labeled by $\rho=\pm1$. For weak strengths of ${\cal B}$, the right $\rho=1$ (left $\rho=-1$) mover is described by the effective Hamiltonian:
\begin{align}
\underline{\hat{{\cal H}}}_{0;\tau_z=1;\rho=\pm1}^{\Delta=0,M_y}(\bm{k})\approx\rho\sqrt{1-(\upsilon\hbar k_y/|\mu|)^2}\left(\upsilon\hbar k_x-M_y-\rho|\mu|\sqrt{1-(\upsilon\hbar k_y/|\mu|)^2}\right)-\frac{\rho\upsilon\hbar{\cal B}}{2|\mu|\sqrt{1-(\upsilon\hbar k_y/|\mu|)^2}}\,.
\end{align}

\noi From the above we can immediately obtain the induced spin density $S_y$ when $M_y\rightarrow0$. We have the expression:
\begin{align}
S_y=\sum_{\rho=\pm1}\int_{-|\mu|/\upsilon\hbar}^{+|\mu|/\upsilon\hbar}\frac{dk_y}{2\pi}\int_{-k_c}^{+k_c}\frac{du}{2\pi\upsilon\hbar}\ph\rho\ph\Theta\left\{-\rho u+|\mu|\left[1-\left(\frac{\upsilon\hbar k_y}{|\mu|}\right)^2\right]+\frac{\rho\upsilon\hbar{\cal B}}{2|\mu|\sqrt{1-(\upsilon\hbar k_y/|\mu|)^2}}\right\}\,,
\end{align}

\noi where $u=\sqrt{1-(\upsilon\hbar k_y/|\mu|)^2}\upsilon \hbar k_x$ and $k_c$ denotes a cutoff wavenumber which controls the validity of the linearization of the energy spectrum about the Fermi points. $S_y$ is zero for ${\cal B}=0$ and at first order in ${\cal B}$, we find:
\begin{align}
S_y=\frac{{\cal B}}{4\pi\upsilon\hbar}\int_{-|\mu|/\upsilon\hbar}^{+|\mu|/\upsilon\hbar}\frac{dk_y}{\pi}\frac{\upsilon\hbar/|\mu|}{\sqrt{1-(\upsilon\hbar k_y/|\mu|)^2}}\frac{1}{2}\sum_{\rho=\pm1}\int_{-k_c}^{+k_c}du\ph \delta\left\{-\rho u+|\mu|\left[1-\left(\upsilon\hbar k_y/|\mu|\right)^2\right]\right\}.
\end{align}

\noi The last integral in the above expression yields unity for each mover, hence leading to the simple expression:
\begin{align}
S_y=\frac{{\cal B}}{4\pi\upsilon\hbar}\int_{-1}^{+1}\frac{d\xi}{\pi}\frac{1}{\sqrt{1-\xi^2}}=\frac{\partial_yM_z}{4\pi\upsilon\hbar}\,.
\end{align}

\noi The above yields the correct expression for $g_{{\color{black}\rm soc}}^{\rm TI}(\Delta=0)$ given that $\mu>0$. Repeating the above process for $\mu<0$, allows us to recover the previous result for both signs of the chemical potential $\mu$.

\subsection{Influence of a Quadratic Dispersion and Warping in the Non-Superconducting Phase}\label{app:Dev}

We obtain $g_{\rm soc}$ in the presence of the additional perturbations using Eq.~\eqref{eq:gspinApp} for the electron part of the Hamiltonian $\hat{{\cal H}}_{0;\tau_z=1}^{\Delta=0}(\bm{k})=(\hbar\bm{k})^2/2m+\upsilon\hbar(k_x\sigma_y-k_y\sigma_x)+\gamma k_x(k_x^2-3k_y^2)\sigma_z-\mu$. After evaluating the various traces we obtain:
\bea
&&g_{\rm soc}^{\rm TI}(\Delta=0)=\int_0^{\infty}\frac{dk}{2\pi}\int_{-\infty}^{+\infty}\frac{d\epsilon}{\pi i}\ph\upsilon\hbar k\left\{\frac{\epsilon-i\mu}{\big[(\epsilon-i\mu)^2+(\upsilon\hbar k)^2\big]^2}-\big(\gamma k^3\big)^2\frac{\epsilon-i\mu}{\big[(\epsilon-i\mu)^2+(\upsilon\hbar k)^2\big]^3}\right\}\no\\
&&-\int_0^{\infty}\frac{dk}{\pi}\int_{-\infty}^{+\infty}\frac{d\epsilon}{\pi}\ph\frac{(\upsilon\hbar k)^3}{m\upsilon^2}\frac{\big(\epsilon-i\mu\big)^2}{\big[(\epsilon-i\mu)^2+(\upsilon\hbar k)^2\big]^3}
+\int_0^{\infty}\frac{dk}{2\pi}\int_{-\infty}^{+\infty}\frac{d\epsilon}{2\pi}\ph\frac{(\upsilon\hbar k)^5}{\big(m\upsilon^2\big)^2}\frac{i\big(\epsilon-i\mu\big)\big[5\big(\epsilon-i\mu\big)^2-(\upsilon\hbar k)^2\big]}{\big[(\epsilon-i\mu)^2+(\upsilon\hbar k)^2\big]^4}\no
\eea

\noi where we have already expanded up to second order in $1/m$ and $\gamma$. One confirms that the contribution of the first term yields ${\rm sgn}(\mu)/4\pi\upsilon\hbar$ and thus leads to the quantization of $g_{{\color{black}\rm soc}}^{\rm TI}$ for $\gamma=0$ and $m\rightarrow\infty$. We focus on the contribution of the remaining terms, which we re-express as follows (we replace $\infty$ by an energy cutoff $\Lambda>0$ for convenience):
\bea
&&g_{{\color{black}\rm soc}}^{\rm TI}(\Delta=0)=\frac{{\rm sgn}(\mu)}{4\pi\upsilon\hbar}+\frac{1}{4\pi\upsilon\hbar}\frac{\gamma^2/2}{\big(\upsilon\hbar\big)^6}\frac{\partial}{\partial\mu}\int_{\Lambda}^0du\ph u^6\frac{\partial}{\partial u}\int_{-\infty}^{+\infty}\frac{d\epsilon}{2\pi}\frac{1}{(\epsilon-i\mu)^2+u^2}\no\\
&&\qquad\quad\phd\phd-\frac{1}{\pi\upsilon\hbar}\frac{1}{m\upsilon^2}\int^0_{\Lambda}du\ph u^2\frac{\partial}{\partial u}\int_{-\infty}^{+\infty}\frac{d\epsilon}{2\pi}\frac{1}{(\epsilon-i\mu)^2+u^2}+\frac{1}{2\pi\upsilon\hbar}\frac{1}{m\upsilon^2}\int^0_{\Lambda}du\ph u^4\frac{\partial}{\partial u}\int_{-\infty}^{+\infty}\frac{d\epsilon}{2\pi}\frac{1}{\big[(\epsilon-i\mu)^2+u^2\big]^2}\no\\
&&+\frac{1}{16\pi\upsilon\hbar}\frac{5}{\big(m\upsilon^2\big)^2}\frac{\partial}{\partial\mu}\int^0_\Lambda du\ph u^4\frac{\partial}{\partial u}\int_{-\infty}^{+\infty}\frac{d\epsilon}{2\pi}\frac{1}{(\epsilon-i\mu)^2+u^2}
-\frac{1}{8\pi\upsilon\hbar}\frac{1}{\big(m\upsilon^2\big)^2}\frac{\partial}{\partial\mu}\int^0_\Lambda du\ph u^6\frac{\partial}{\partial u}\int_{-\infty}^{+\infty}\frac{d\epsilon}{2\pi}\frac{1}{\big[(\epsilon-i\mu)^2+u^2\big]^2}
\no\\\no\\
%\no\\
%
&&\phantom{g_{{\color{black}\rm soc}}^{\rm TI}(\Delta=0)}=\frac{{\rm sgn}(\mu)}{4\pi\upsilon\hbar}+\frac{1}{4\pi\upsilon\hbar}\frac{\gamma^2/2}{\big(\upsilon\hbar\big)^6}\frac{\partial}{\partial\mu}\int_{\Lambda}^0du\ph u^6\frac{\partial}{\partial u}\frac{\Theta\big(u-|\mu|\big)}{2u}-\frac{1}{\pi\upsilon\hbar}\frac{1}{m\upsilon^2}\int^0_{\Lambda}du\ph u^2\frac{\partial}{\partial u}\frac{\Theta\big(u-|\mu|\big)}{2u}\no\\
&&\qquad\qquad\quad\phd-\frac{1}{8\pi\upsilon\hbar}\frac{1}{m\upsilon^2}\int^0_{\Lambda}du\ph u^4\frac{\partial}{\partial u}\left\{\frac{1}{u^2}\left[\delta(u-|\mu|)-\frac{\Theta(u-|\mu|)}{u}\right]\right\}
+\frac{1}{16\pi\upsilon\hbar}\frac{5}{\big(m\upsilon^2\big)^2}\frac{\partial}{\partial\mu}\int^0_\Lambda du\ph u^4\frac{\partial}{\partial u}\frac{\Theta\big(u-|\mu|\big)}{2u}\no\\
&&\qquad\qquad\quad\phd
+\frac{1}{32\pi\upsilon\hbar}\frac{1}{\big(m\upsilon^2\big)^2}\frac{\partial}{\partial\mu}\int^0_\Lambda du\ph u^6\frac{\partial}{\partial u}\left\{\frac{1}{u^2}\left[\delta(u-|\mu|)-\frac{\Theta(u-|\mu|)}{u}\right]\right\}.\no
\eea

\noi Carrying out the integrations leads to the following result:
\begin{align}
g_{{\color{black}\rm soc}}^{\rm TI}(\Delta=0)=\frac{{\rm sgn}(\mu)}{4\pi\upsilon\hbar}\left\{1+\frac{1}{2}\left(\frac{\mu}{m\upsilon^2}\right)^2-\frac{3}{2}\left[\frac{\gamma\mu^2}{\big(\upsilon\hbar\big)^3}\right]^2\right\}
-\frac{1}{8\pi\upsilon\hbar}\frac{\Lambda}{m\upsilon^2}\,.
\label{eq:gspinCorrected}
\end{align}

The above is obtained after omitting terms $\propto \delta(\Lambda-|\mu|)$ and $\partial_\mu\Theta(\Lambda-|\mu|)$, since we assume that $\Lambda\gg|\mu|$. Following the same spirit as in Sec.~\ref{sec:ZeemanEffects}, we drop the contribution $\propto\Lambda$, since we are interested in the response which originates from energies near the Fermi energy $|\mu|\ll\Lambda$. Here, one can also invoke that $m\upsilon^2\gg\Lambda$, which allows dropping the last term. Therefore, the above analysis implies that the quantization found for $g_{\rm soc}^{\rm TI}(\Delta=0)$ for the pristine Dirac cone surface states is protected against adding warping and a quadratic dispersion of a weak strength. This is because the modification is of quadratic order in $1/m$ and $\gamma$. Hence, the magnitude of such corrections relative to the quantized value of $g_{{\color{black}\rm soc}}^{\rm TI}(\Delta=0)$ introduced by the two kinds of perturbations depends on the value of the chemical potential, and how close this is to the Dirac point. As it is typical for similar topological semimetals, the various anomalous properties prevail as long as the chemical potential is tuned near the band touching point.

We now proceed with examining the fate of $\chi^{\rm TI}(\Delta=0)$ when these two kinds of deviations from the pristine Dirac cone structure are present. In the same spirit with the previous paragraphs, we proceed by obtaining the modified coefficient from Eq.~\eqref{eq:chi}. However, the respective polarization tensor elements need to be modified according to:
\bea
\Pi_{M_zA_x}(\bm{q})&=&\int\frac{d\bm{k}}{(2\pi)^2}T\sum_{ik_n}\frac{1}{2}{\rm Tr}\left\{\sigma_z\hat{\cal G}_0'(\bm{k}+\bm{q},ik_n)(+e\upsilon\sigma_y)\hat{\cal G}_0'(\bm{k},ik_n)\right\}\no\\
&+&\int\frac{d\bm{k}}{(2\pi)^2}T\sum_{ik_n}\frac{1}{2}{\rm Tr}\left\{\sigma_z\hat{\cal G}_0'(\bm{k}+\bm{q},ik_n)\frac{e}{\hbar}\left\{\frac{\hbar^2}{m}\left(k_x+\frac{q_x}{2}\right)+3\gamma\left[k_x^2-k_y^2+q_x\left(k_x+\frac{q_x}{3}\right)\right]\sigma_z\right\}\hat{\cal G}_0'(\bm{k},ik_n)\right\}\no\\
\Pi_{M_zA_y}(\bm{q})&=&\int\frac{d\bm{k}}{(2\pi)^2}T\sum_{ik_n}\frac{1}{2}{\rm Tr}\left\{\sigma_z\hat{\cal G}_0'(\bm{k}+\bm{q},ik_n)(-e\upsilon\sigma_x)\hat{\cal G}_0'(\bm{k},ik_n)\right\}\no\\
&+&\int\frac{d\bm{k}}{(2\pi)^2}T\sum_{ik_n}\frac{1}{2}{\rm Tr}\left\{\sigma_z\hat{\cal G}_0'(\bm{k}+\bm{q},ik_n)\frac{e}{\hbar}\left\{\frac{\hbar^2}{m}\left(k_y+\frac{q_y}{2}\right)-3\gamma\big(k_x+q_x\big)\big(2k_y+q_y\big)\sigma_z\right\}\hat{\cal G}_0'(\bm{k},ik_n)\right\}\no
\eea

\noi where the above Green functions $\hat{\cal G}_0'$ include the perturbation terms $\propto \gamma,\,1/m$. The above are separated into two parts, one per row. The parts contained in the first rows give rise to a contribution which satisfies $\chi^{\rm TI}(\Delta=0)=e\upsilon g_{{\color{black}\rm soc}}^{\rm TI}(\Delta=0)$ even when $\gamma$ and $1/m$ are generally nonzero. Hence, the corrections due to the two pertubations introduced to $\chi^{\rm TI}(\Delta=0)$ arising from the first rows are immediately obtained using Eq.~\eqref{eq:gspinCorrected}. However, $\chi^{\rm TI}(\Delta=0)$ takes an additional contribution originating from the second rows. These result from the conservation of electric charge which leads to the modified vertices. Concomitantly this generally implies the inequivalence $\chi^{\rm TI}(\Delta=0)\neq e\upsilon g_{{\color{black}\rm soc}}^{\rm TI}(\Delta=0)$ for the final expressions. Direct evaluation of Eq.~\eqref{eq:chi} yields the expression:
\bea
\chi^{\rm TI}(\Delta=0)&=&e\upsilon g_{{\color{black}\rm soc}}^{\rm TI}(\Delta=0)+\frac{e}{4\pi\hbar}\int^\Lambda_0du\left(\frac{u^2}{m\upsilon^2}\right)\left[1-\frac{1}{2}\left(\frac{u^2}{m\upsilon^2}\right)\frac{\partial}{\partial\mu}\right]\frac{\partial}{\partial u}\int_{-\infty}^{+\infty}\frac{d\epsilon}{2\pi}\ph\frac{1}{(\epsilon-i\mu)^2+u^2}\no\\
%
% &=&e\upsilon g_{{\color{black}\rm soc}}^{\rm TI}(\Delta=0)+\frac{e}{4\pi\hbar}\left[{\rm sgn}(\mu)\frac{\mu}{m\upsilon^2}-\frac{\Lambda}{2m\upsilon^2}-{\rm sgn}(\mu)\left(\frac{\mu}{m\upsilon^2}\right)^2\right]\no\\
%
&=&\frac{e}{4\pi\hbar}{\rm sgn}(\mu)\left\{1+\frac{\mu}{m\upsilon^2}-\frac{1}{2}\left(\frac{\mu}{m\upsilon^2}\right)^2-\frac{3}{2}\left[\frac{\gamma\mu^2}{\big(\upsilon\hbar\big)^3}\right]^2\right\}
-\frac{e}{4\pi\hbar}\frac{\Lambda}{m\upsilon^2}\,.
\label{eq:chiCorrected}
\eea

\noi Once again, the term $\propto\Lambda$ needs to be dropped. Notably, besides the part of $\chi^{\rm TI}(\Delta=0)$ which is proportional to the modified $g_{{\color{black}\rm soc}}^{\rm TI}(\Delta=0)$, the additional term affects $\chi^{\rm TI}(\Delta=0)$ at first order with in $1/m$, due to the vertex correction $\propto\bm{k}/m$. This implies that $\chi^{\rm TI}(\Delta=0)$ is less protected than $g_{{\color{black}\rm soc}}^{\rm TI}(\Delta=0)$.

\subsection{Investigation of Deviations from the Pristine Dirac Cone via the Thery of Orbital Magnetization}\label{app:Dev2}

For completeness we obtain once again the results of Appendix~\ref{app:Dev} using the theory of orbital magnetization. We extend the approach of Sec.~\ref{sec:OM} to include hexagonal warping and a quadratic kinetic energy term. For the two band model of interest, the uniform orbital magnetization $T=0$ now becomes:
\begin{align}
{\cal M}_z=\frac{e}{4\pi\hbar}\int\frac{d\bm{k}}{2\pi}\ph2\mu(\bm{k})\Omega(\bm{k})\Big\{\Theta\big[\mu(\bm{k})+E(\bm{k})\big]-\Theta\big[\mu(\bm{k})-E(\bm{k})\big]\Big\}
\end{align}

\noi where we have introduced the $\bm{k}$-dependent chemical potential $\mu(\bm{k})=\mu-(\hbar\bm{k})^2/2m$. Furthermore, we parametrized the ensuing Hamiltonian $\hat{{\cal H}}_{0;\tau_z=1}^{\Delta=0,M_z\neq0}(\bm{k})=(\hbar\bm{k})^2/2m+\upsilon\hbar(k_x\sigma_y-k_y\sigma_x)+\big[\gamma k_x(k_x^2-3k_y^2)-M_z\big]\sigma_z-\mu$ according to $\hat{{\cal H}}_{0;\tau_z=1}^{\Delta=0,M_z\neq0}(\bm{k})\equiv\bm{d}(\bm{k})\cdot\bm{\sigma}-\mu(\bm{k})$, where $\bm{d}(\bm{k})=\big(-\upsilon\hbar k_y,\upsilon\hbar k_x,\gamma k_x(k_x^2-3k_y^2)-M_z\big)$. The latter vector possesses the modulus $E(\bm{k})\equiv|\bm{d}(\bm{k})|$. Here, $\Omega(\bm{k})$ corresponds to the Berry curvature of the valence band, and is given as $\Omega(\bm{k})=\frac{1}{2}\hat{\bm{d}}(\bm{k})\cdot\big[\partial_{k_x}\hat{\bm{d}}(\bm{k})\times\partial_{k_y}\hat{\bm{d}}(\bm{k})\big]$, where $\hat{\bm{d}}(\bm{k})=\bm{d}(\bm{k})/E(\bm{k})$. Using the above expression, it is straightforward to obtain the orbital magnetization up to second order in $1/m$ and $\gamma$. By means of a Taylor expansion, we have:
\bea
{\cal M}_z
=\frac{e}{4\pi\hbar}\left\{\sum_{s=0,1,2}\frac{(-1)^s}{s!\big[2m\upsilon^2\big]^s}\frac{\partial^s}{\partial\mu^s}\int\frac{d\bm{k}}{2\pi}\big(\upsilon\hbar\bm{k}\big)^{2s}m(k)-\frac{\gamma^2}{2}\int\frac{d\bm{k}}{2\pi}\big[k_x(k_x^2-3k_y^2)\big]^2\frac{\partial}{\partial M_z}\left[6\frac{m(k)}{M_z}-\frac{\partial m(k)}{\partial M_z}\right]\right\}\no
\eea

\noi where we retained terms up to second order in $1/m$ and $\gamma$. For compactness, in the above we introduced the function $m(k)=2\mu\Omega_{\gamma=0}(k)\Theta\big[E_{\gamma=0}(k)-|\mu|\big]$ with $\Omega_{\gamma=0}(k)=-(\upsilon\hbar)^2M_z/\big[2E_{\gamma=0}^3(k)\big]$, and $E_{\gamma=0}(k)=\sqrt{(\upsilon\hbar k)^2+M_z^2}$. The desired coefficient at zeroth order in $M_z$, is obtained via the expression $\chi^{\rm TI}(\Delta=0)=-\left.\partial{\cal M}_z/\partial M_z\right|_{M_z=0}$. The result obtained by means of evaluating the above expression is in perfect agreement with the part of Eq.~\eqref{eq:chiCorrected} which is independent of $\Lambda$. We note that there is some discrepancy when it comes to the terms $\propto\Lambda$, which are nevertheless supposed to be neglected since they are spurious and unphysical.

\section{Evaluation of the Interconversion Coefficients: Rashba Metal}\label{app:Appendix3}

We now discuss the evaluation of the interconversion coefficients in the case of a Rashba metal. We employ Eqs.~\eqref{eq:gspin}-\eqref{eq:ChiExpression} of the main text. To facilitate the calculations we introduce polar coordinates, i.e., $k_x=k\cos\varphi$ and $k_y=k\sin\varphi$, where $k=\sqrt{k_x^2+k_y^2}$ and $\tan\varphi=k_y/k_y$. This allows us to re-express the matrix Green function $\hat{\cal G}_0(\bm{k},\epsilon)$ as follows:
\bea
\hat{\cal G}_0^{-1}(k,\varphi,\epsilon)=e^{-i\varphi\sigma_z/2}\hat{g}^{-1}(k,\epsilon)e^{i\varphi\sigma_z/2}
\eea

\noi where $\hat{g}^{-1}(k,\epsilon)=i\epsilon-\tau_z\big[\hbar^2(k-k_F)^2/2m+\upsilon\hbar k\sigma_y\big]-\Delta\tau_x$. Using the above, we find that:
\bea
k_x\frac{\partial\hat{\cal G}_0(\bm{k},\epsilon)}{\partial k_y}-k_y\frac{\partial\hat{\cal G}_0(\bm{k},\epsilon)}{\partial k_x}=\frac{\partial\hat{\cal G}_0(k,\varphi,\epsilon)}{\partial\varphi}=e^{-i\varphi\sigma_z/2}\frac{[\sigma_z,\hat{g}(k,\epsilon)]}{2i}e^{i\varphi\sigma_z/2}\no
\eea

\noi along with the relation:
\bea
\frac{\partial\hat{\cal G}_0(\bm{k},\epsilon)}{\partial k_x}\sigma_x+\frac{\partial\hat{\cal G}_0(\bm{k},\epsilon)}{\partial k_y}\sigma_y=e^{-i\varphi\sigma_z/2}\left\{\frac{\partial\hat{g}(k,\epsilon)}{\partial k}\sigma_x+\frac{1}{k}\frac{[\sigma_z,\hat{g}(k,\epsilon)]}{2i}\sigma_y\right\}e^{i\varphi\sigma_z/2}.
\eea

\noi By exploiting the above results, we find:
\bea
\chi^{\rm M}&=&\frac{e\upsilon_F}{2i}\int_0^\infty\frac{dk}{2\pi}\frac{k}{k_F}\int_{-\infty}^{+\infty}\frac{d\epsilon}{2\pi}\frac{1}{2}{\rm Tr}\left\{\sigma_z\frac{[\sigma_z,\hat{g}(k,\epsilon)]}{2i}\hat{g}(k,\epsilon)\right\}
+\frac{e\upsilon}{2i}\int_0^\infty\frac{kdk}{2\pi}\int_{-\infty}^{+\infty}\frac{d\epsilon}{2\pi}\frac{1}{2}{\rm Tr}\left[\sigma_z\frac{\partial\hat{g}(k,\epsilon)}{\partial k}\sigma_x\hat{g}(k,\epsilon)\right]\no\\
&+&\frac{e\upsilon}{2i}\int_0^\infty\frac{dk}{2\pi}\int_{-\infty}^{+\infty}\frac{d\epsilon}{2\pi}\frac{1}{2}{\rm Tr}\left\{\sigma_z\frac{[\sigma_z,\hat{g}(k,\epsilon)]}{2i}\sigma_y\hat{g}(k,\epsilon)\right\}\,.\no
\eea

At this point, we proceed by implementing the so-called quasiclassical approximation. Thus, by assuming that the Fermi energy $E_F=\hbar^2k_F^2/2m$ is the largest energy scale (essentially infinite), we approximate $\xi=\hbar^2(k-k_F)^2/2m$ as $\xi\approx\hbar\upsilon_F(k-k_F)$ and $\hbar\upsilon k\approx \hbar\upsilon k_F\equiv E_{\rm soc}$. Employing the quasiclassical approximation and tracing over spins yields:
\bea
\chi^{\rm M}&=&-\frac{e}{4\pi\hbar}\int_{-E_F}^{+\infty}d\xi\int_{-\infty}^{+\infty}\frac{d\epsilon}{2\pi}\sum_{\ell,s}^\pm\frac{\ell\cdot s}{4}{\rm Tr}_\tau\big[\hat{g}_\ell(\xi,\epsilon)\hat{g}_s(\xi,\epsilon)\big]\no\\
&&-\frac{e}{4\pi\hbar}E_{\rm soc}\int_{-E_F}^{+\infty}d\xi\int_{-\infty}^{+\infty}\frac{d\epsilon}{2\pi}\sum_{\ell,s}^\pm\frac{\ell-s}{4}{\rm Tr}_\tau\left[\frac{\partial\hat{g}_\ell(\xi,\epsilon)}{\partial\xi}\hat{g}_s(\xi,\epsilon)\right]\no\\
&&-\frac{e}{4\pi\hbar}\frac{\upsilon}{\upsilon_F}\int_{-E_F}^{+\infty}d\xi\int_{-\infty}^{+\infty}\frac{d\epsilon}{2\pi}\sum_{\ell,s}^\pm\frac{\ell}{4}{\rm Tr}_\tau\big[\hat{g}_\ell(\xi,\epsilon)\hat{g}_s(\xi,\epsilon)\big]\,,\no
\eea

\noi where we introduced the quasiclassical Green functions with $\hat{g}^{-1}_\pm(k,\epsilon)=i\epsilon-\tau_z\big[\hbar\upsilon_F(k-k_F)\pm E_{\rm soc}\big]-\Delta\tau_x$. In the above expression the first term contributes to $g_{\rm orb}$ and the last two to $g_{{\color{black}\rm soc}}$. However, in the quasiclassical limit the third term drops out. To continue, we further take the limit $E_F\rightarrow\infty$ and we conclude with the following expressions:
\bea
g_{\rm orb}^{\rm M,intra}&=&-\frac{1}{4\pi\hbar\upsilon_F}\int_{-\infty}^{+\infty}d\xi\int_{-\infty}^{+\infty}\frac{d\epsilon}{2\pi}\sum_{\ell=\pm}\frac{1}{4}{\rm Tr}_\tau\Big[\hat{g}_\ell^2(\xi,\epsilon)\Big]\,,\no\\
g_{\rm orb}^{\rm M, inter}&=&+\frac{1}{4\pi\hbar\upsilon_F}\int_{-\infty}^{+\infty}d\xi\int_{-\infty}^{+\infty}\frac{d\epsilon}{2\pi}\sum_{\ell=\pm}\frac{1}{2}{\rm Tr}_\tau\big[\hat{g}_+(\xi,\epsilon)\hat{g}_-(\xi,\epsilon)\big]\,,\no\\
g_{{\color{black}\rm soc}}^{\rm M}&=&-\frac{E_{\rm soc}}{4\pi\hbar\upsilon_F}\int_{-\infty}^{+\infty}d\xi\int_{-\infty}^{+\infty}\frac{d\epsilon}{2\pi}\sum_{\ell=\pm}\frac{\ell}{2}{\rm Tr}_\tau\left[\frac{\partial\hat{g}_\ell(\xi,\epsilon)}{\partial\xi}\hat{g}_{-\ell}(\xi,\epsilon)\right]\,.\no
\eea

The evaluation of the coefficients $g_{\rm orb}^{\rm M,inter}$ and $g_{{\color{black}\rm soc}}^{\rm M}$ is tedious but straightforward. Therefore, in the remainder, we focus on the derivation of the coefficient $g_{\rm orb}^{\rm M,intra}$. Since this coefficient admits only intraband contributions, it is preferrable to momentarily switch back to the Matsubara formalism. Carrying out the Matsubara summation yields:
\bea
g_{\rm orb}^{\rm M,intra}=-\frac{1}{4\pi\hbar\upsilon_F}\int_{-\infty}^{+\infty}d\xi\sum_{\ell=\pm}\frac{1}{2}n_F'\big[E_\ell(\xi)\big],\no
\eea

\noi where we introduced the energy dispersions of the two helical bands $E_\pm(\xi)=\sqrt{(\xi\pm E_{\rm soc})^2+\Delta^2}$. $n_F(E)$ denotes the Fermi-Dirac distribution evaluated at energy $E$, while $n_F'(E)=dn_F(E)/dE$. At zero tem\-pe\-ra\-tu\-re $n_F'(E)=-\delta(E)$, with the latter function denoting the Dirac delta function. The intraband nature of this contribution implies that it is sensitive to the presence of disorder and nonzero temperature. We examine the effects of disorder in a phenomenological and qualitative fashion by ``broadening'' the Fermi-Dirac function according to~\cite{SteffensenNematic}:
\bea
-n_F'\big[E_\ell(\xi)\big]=\int_{-\infty}^{+\infty}\frac{d\omega}{2\pi}\frac{1}{\tau_\ell(\xi)}\frac{-n_F'\big(\omega\big)}{\big[\omega-E_\ell(\xi)\big]^2+\left(\frac{1}{2\tau_\ell(\xi)}\right)^2}
\eea

\noi where $\tau_\ell(\xi)$ define effective intra-band relaxation times. In the remainder, we consider them to be energy independent, in order to facilitate the discussion, and express them as $\tau_\ell=1/(2{\color{black}\Gamma}_\ell)$. By restricting to $T=0$, we have the expression:
\begin{align}
-n_F'\big[E_\ell(\xi)\big]=\int_{-\infty}^{+\infty}\frac{d\omega}{2\pi}\frac{1}{\tau_\ell}\frac{\delta(\omega)}{\big[\omega-E_\ell(\xi)\big]^2+\left(\frac{1}{2\tau_\ell}\right)^2}=\frac{1}{\pi}\frac{{\color{black}\Gamma}_\ell}{E_\ell^2(\xi)+{\color{black}\Gamma}_\ell^2}\,.
\end{align}

\noi Therefore, the interconversion coefficient now reads as:
\begin{align}
g_{\rm orb}^{\rm M,intra}=\frac{1}{4\pi\hbar\upsilon_F}\sum_{\ell=\pm}\frac{1}{2}\int_{-\infty}^{+\infty}\frac{d\xi}{\pi}\frac{{\color{black}\Gamma}_\ell}{\big(\xi+\ell E_{\rm soc}\big)^2+\Delta^2+{\color{black}\Gamma}_\ell^2}
\equiv\frac{1}{4\pi\hbar\upsilon_F}\sum_{\ell=\pm}\frac{1}{2}\int_{-\infty}^{+\infty}\frac{d\xi}{\pi}\frac{{\color{black}\Gamma}_\ell}{\xi^2+\Delta^2+{\color{black}\Gamma}_\ell^2}.
\end{align}

\noi Carrying out the integration over $\xi$ directly yields the result of Eq.~\eqref{eq:intraband} in the main text.

{\color{black}

\section{Out-of-plane Spin Susceptibility for a Rashba Metal}\label{app:Appendix4}

We now proceed with the evaluation of the out-of-plane spin susceptibility in the case of a Rashba metal in the quasiclassical limit. It is most convenient to obtain $\chi_\perp^{\rm spin}$ by including a uniform out-of-plane magnetization $M_z$ to the Hamiltonian of the Rashba superconductor, which results in the expression:
\begin{align}
\hat{\cal H}(\bm{k})=\tau_z\big[\varepsilon(k)+\upsilon\hbar (k_x\sigma_y-k_y\sigma_x)\big]+\Delta\tau_x-M_z\sigma_z,\no
\end{align}

\noi where $\varepsilon(k)=(\hbar k)^2/2m-\mu$ with $k=|\bm{k}|$. The above Hamiltonian is dictated by the eigenergies $\pm E_\pm(k)$ given by:
\begin{align}
E_\pm(k)=\sqrt{\varepsilon^2(k)+(\upsilon\hbar k)^2+M_z^2+\Delta^2\pm2\sqrt{{\cal R}(k)}}\no
\end{align}

\noi where we introduced the quantity:
\begin{align}
{\cal R}(k)=(\upsilon\hbar k)^2\varepsilon^2(k)+M_z^2\varepsilon^2(k)+M_z^2\Delta^2\,.
\end{align}

\noi The susceptibility is obtained from the definition:
\begin{align}
\chi_\perp=-\left.\frac{d^2E_{\rm gs}}{dM_z^2}\right|_{M_z=0}\,,
\end{align}

\noi where $E_{\rm gs}$ is the ground state energy of the system per area. Since the spectrum depends only on $k$, we find that:
\begin{align}
E_{\rm gs}=-\frac{1}{2}\sum_{s=\pm}\int_0^\infty\frac{kdk}{2\pi}E_s(k)\,.
\end{align}

We adopt a quasiclassical approach and set $\varepsilon(k)\approx\hbar\upsilon_F(k-k_F)$, $E_{\rm soc}=\upsilon\hbar k_F$. By further taking the limit $E_F\rightarrow\infty$, the ground state energy becomes:
\begin{align}
E_{\rm gs}=-\frac{\nu_F}{2}\sum_{s=\pm}\int_{-\infty}^{+\infty}d\varepsilon\ph\sqrt{\varepsilon^2+E_{\rm soc}^2+M_z^2+\Delta^2+2s\sqrt{{\cal R}(\varepsilon)}},
\end{align}

\noi where $\nu_F=m/2\pi\hbar^2$ and we set ${\cal R}(\varepsilon)=(E_{\rm soc}^2+M_z^2)\varepsilon^2+M_z^2\Delta^2$. By directly evaluating the above integral we obtain the susceptiblity expression in Eq.~\eqref{eq:SpinSuscMetal}.}

\end{widetext}

\end{document}